%% file: doc.tex
\documentclass[a4paper,twocolumn,10pt]{quantumarticle}
\pdfoutput=1
\usepackage[utf8]{inputenc}
\usepackage[english]{babel}
\usepackage[T1]{fontenc}
\usepackage{amsmath}
\usepackage{amssymb}
\usepackage{braket}
\usepackage{tikz}
\usepackage{multirow}
\usepackage{hyperref}
\usepackage{subcaption}
\usepackage{algorithm2e}
\usepackage{listings}
\def\HiLi{\leavevmode\rlap{\hbox to \hsize{\color{blue!20}\leaders\hrule height .8\baselineskip depth .5ex\hfill}}}

\newtheorem{example}{Example}

\begin{document}

\title{Multiple Query Optimization using a Hybrid Approach of Classical and Quantum Computing}

\author{Tobias Fankhauser}
\email{tobias7402@gmail.com}
\affiliation{Zurich University of Applied Sciences, Winterthur, Switzerland}
\orcid{}
\author{Marc E. Solèr}
\email{marcesoler@sns.network}
\affiliation{Zurich University of Applied Sciences, Winterthur, Switzerland}
\author{Rudolf M. Füchslin}
\affiliation{Zurich University of Applied Sciences, Winterthur, Switzerland}
\affiliation{European Centre for Living Technology (ECLT), Ca' Bottacin, Dorsoduro 3911, 30123 - Venice, Italy}
\orcid{0000-0002-9296-5314}
\email{rudolf.fuechslin@zhaw.ch}
\author{Kurt Stockinger}
\affiliation{Zurich University of Applied Sciences, Winterthur, Switzerland}
\orcid{0000-0003-4034-4812}
\email{kurt.stockinger@zhaw.ch}

\begin{abstract}
Quantum computing promises to solve difficult optimization problems in chemistry, physics and mathematics more efficiently than classical computers, but requires fault-tolerant quantum computers with millions of qubits. To overcome errors introduced by today's quantum computers, hybrid algorithms combining classical and quantum computers are used.

In this paper we tackle the multiple query optimization problem (MQO) which is an important NP-hard problem in the area of data-intensive problems. We propose a \textit{novel hybrid classical-quantum algorithm to solve the MQO on a gate-based quantum computer}. We perform a detailed experimental evaluation of our algorithm and compare its performance against a competing approach that employs a quantum annealer -- another type of quantum computer. Our experimental results demonstrate that our algorithm currently can only handle small problem sizes due to the limited number of qubits available on a gate-based quantum computer compared to a quantum computer based on quantum annealing. However, our algorithm shows a \textit{qubit efficiency of close to 99\%} which is almost a \textit{factor of 2 higher} compared to the state of the art implementation. Finally, we analyze how our algorithm scales with larger problem sizes and conclude that our approach shows promising results for near-term quantum computers.
\end{abstract}

\section{Introduction}\label{sec:introduction}
In database research, various optimization problems have been formulated since the seventies \cite{selinger1989access, heitz2019join, sellis1988multiple}, with the query optimization problem being a typical NP-hard representative.
These problems, although extensively examined, become even more difficult as data processing becomes more complex \cite{trummer2015dwave}.
Recent approaches to the query optimization problem have also become increasingly sophisticated, now employing deep learning methods \cite{heitz2019join, marcus2019neo}.
Simultaneously, progress in the construction of quantum computers has sparked new interest in applied quantum computing  \cite{bergholm2018pennylane, farhi2014quantum, google_quantum, peruzzo2014variational}, whereas previously, the quantum computing community was mainly concerned with more theoretical studies due to the lack of practical quantum devices \cite{moll2018quantum}.
It is hoped that quantum computers can one day solve complex problems more efficiently than classical computers \cite{moll2018quantum}.

Although quantum computers were first envisioned to simulate quantum systems \cite{feynman1982simulating}, the first two famous use cases were outside of physics. Namely, search in an unstructured list \cite{grover1996fast} and factoring large integers \cite{shor1999polynomial}.

More recently, quantum computers are considered for mathematical optimization \cite{farhi2014quantum}, or simulation of molecules for chemistry \cite{peruzzo2014variational}.
Current quantum computers are still in their infancy and are restricted by the capacity and fault-tolerance \cite{moll2018quantum,google_quantum,zhou2020quantum}, making them more interesting for quantum research than for real-world application. 
This motivated the creation of hybrid classical-quantum algorithms like the Quantum Approximate Optimization Algorithm (QAOA) \cite{farhi2014quantum} that employ a quantum device to explore high-dimensional search space and classical routines to optimize the search procedure.
By only relaying parts of the computation to the quantum computers, these hybrid classical-quantum-algorithms are more resistant to error \cite{farhi2014quantum}, making them more suitable to near-term devices.

The Multiple Query Optimization problem (MQO) is a generalization of the query optimization problem \cite{sellis1988multiple}. MQO is known to be an NP-hard problem \cite{sellis1990nphard}. It has been approached by classical methods such as the shortest-path algorithms \cite{sellis1988multiple}, genetic algorithms \cite{bayir2006genetic} and others.
Recently, a quantum approach was suggested for the MQO by Trummer and Koch \cite{trummer2015dwave}, utilizing a D-Wave quantum annealer \cite{dwave}, a particular type of quantum computer designed to run special optimization problems \cite{trummer2015dwave}.
Our work builds on Trummer and Koch's approach, but utilizes a gate-based quantum computer, a more general type of quantum computer. Gate-based quantum computers can, in principle, run every quantum algorithm \cite{Qiskit,ibmq-manhattan}, while quantum annealers are restricted to specific problems \cite{trummer2015dwave}.
This advantage comes with the cost that gate-based quantum computers currently feature significantly fewer qubits and are more error-prone than quantum annealers.

\subsection{Multiple Query Optimization}
Multiple query optimization (MQO) aims to minimize the total cost of executing a series of queries against a database by exploiting shared intermediate results \cite{sellis1988multiple}.

\begin{figure}[h]
    \centering
    \begin{subfigure}[b]{0.35\textwidth}
        \centering
        \includegraphics[width=0.8\textwidth]{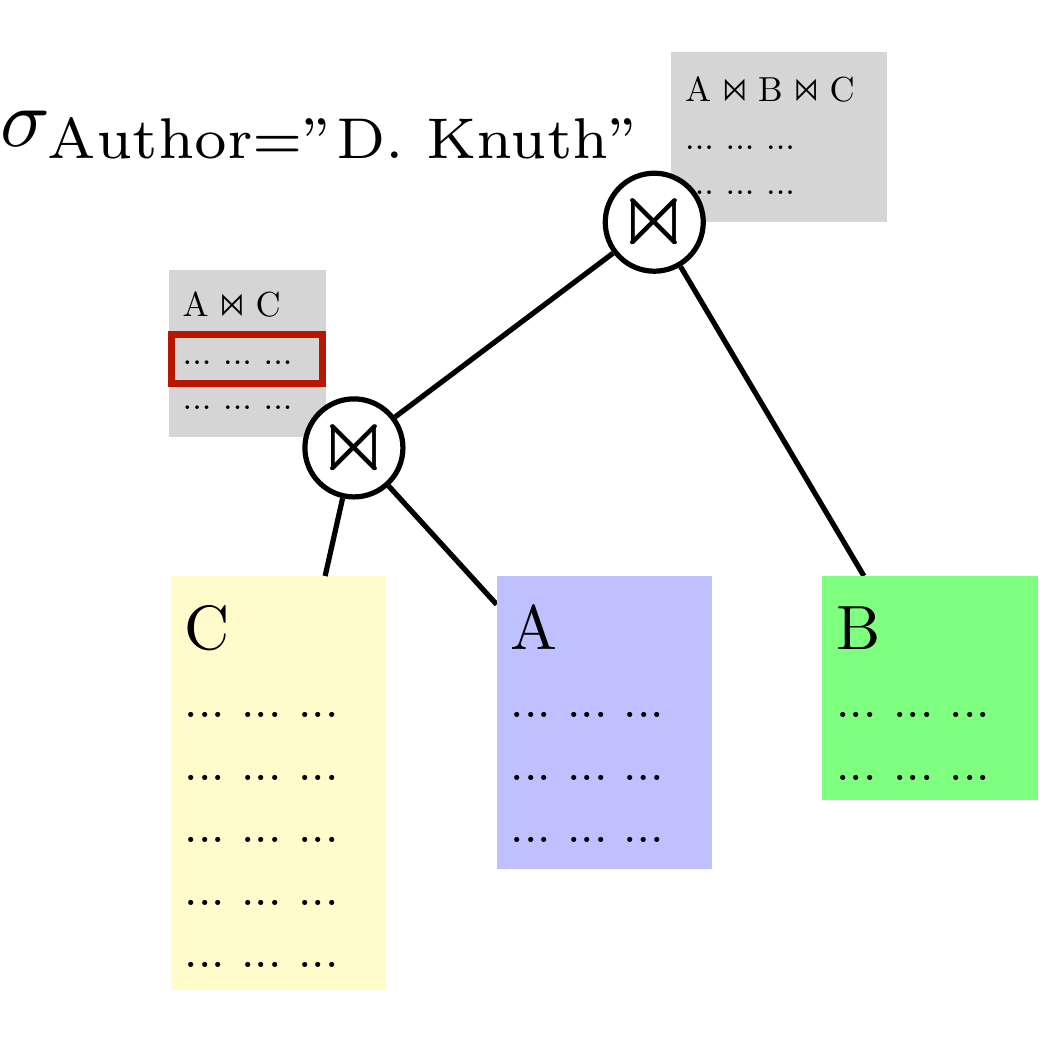}
        \caption{}
    \end{subfigure}
    \begin{subfigure}[b]{0.35\textwidth}
        \centering
        \includegraphics[width=0.8\textwidth]{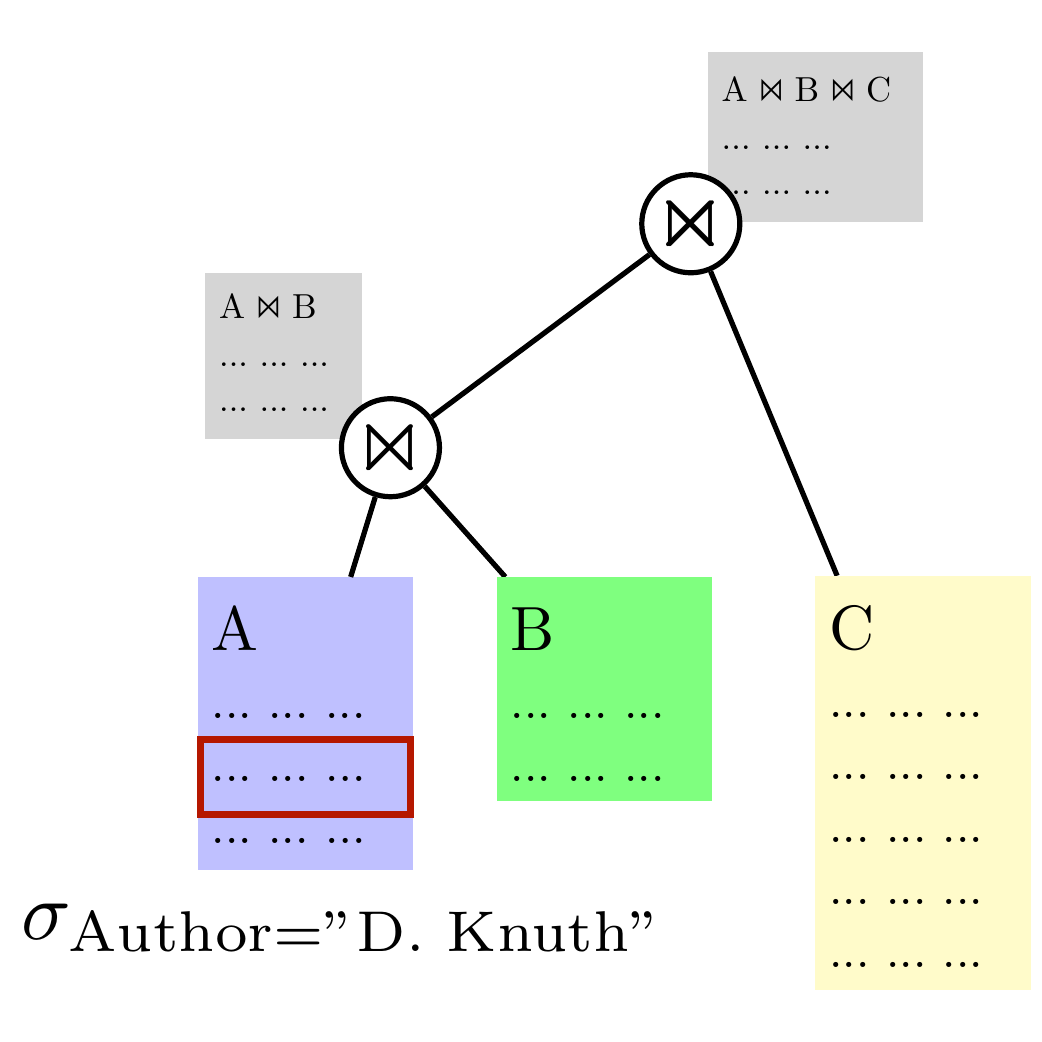}
        \caption{}
    \end{subfigure}
    \caption[Two possible join plans]{Optimizing queries against databases is an NP-hard problem. This figure shows two different query plans, i.e. partial solutions of the combinatorial optimization problem. (a) shows a query plan where table $C$ (largest table) is joined with $A$ (medium table). Afterwards a selection on the author "D. Knuth" is applied on the intermediate result which is then joined with table $B$ (smallest table). Because the selection is done late, the join operation has to be applied on a large result set. (b) shows a more efficient query plan, where the selection is first done on table $A$ (medium table) resulting in a smaller result set -- prior to the join with $B$ (smallest table). The intermediate result is finally joined with table $C$ (largest table).}
    \label{fig:plan_comparison}
\end{figure}

\begin{example}
Consider a database with tables $A, B, \ldots, H$ of different sizes. On these tables, queries can be formulated, such as $\sigma_{\text{Author="D. Knuth"}}(A \Join B \Join C)$. To produce this query, there may exist multiple plans that vary in the order of joins, order of selections and other aspects. For example, selections are usually done before joining tables to reduce the number of operations required for the join. Figure \ref{fig:plan_comparison} shows two of those plans.
\end{example}

Our model considers a series of queries $q_1, \ldots, q_Q$, where each query has a fixed number $P$ of plans, uniquely identified as $p_1, \ldots, p_{PQ}$. For every query, exactly one plan has to be selected. Each plan $p_i$ has associated a cost $c_i$, which our model assumes as given.
Finally, we denote savings (e.g. from shared intermediate results) pairwise between plans $p_i$ and $p_j$ as $s_{i, j}$. Savings can be subtracted from the total cost if both plans from the pair have been selected.
In the below example, the pairs of plans $p_3$, $p_4$ have savings as they can use an intermediate result from sharing a common table $B$.  The goal is to find a selection of plans for all the queries, such that the overall cost is minimized. \\

\begin{centering}
\begin{tabular}{c c c c}
    $A \Join$ \textcolor{blue}{$B$} $\Join C$ & \textcolor{blue}{$B$} $\Join$ \textcolor{violet}{$D$} & $\cdots$ & \textcolor{violet}{$D$} $\Join G \Join H$  \\
    $q_1$ & $q_2$ & $\ldots$ & $q_Q$ \\
    \textcolor{gray}{$p_1$}, \textcolor{gray}{$p_2$}, \textcolor{blue}{$p_3$} & \textcolor{blue}{$p_4$}, \textcolor{gray}{$p_5$}, \textcolor{violet}{$p_6$} & $\ldots$ & \textcolor{violet}{$p_{PQ-2}$}, \textcolor{gray}{$p_{PQ-1}$}, \textcolor{gray}{$p_{PQ}$} \\
    \multicolumn{2}{c}{\textcolor{blue}{$s_{3,4}$}} &  \multicolumn{2}{l}{\textcolor{violet}{$s_{6,PQ-2}$}}\\
\end{tabular}\\
\medskip
\end{centering}

The search space of this problem is significant as there are $P$ possibilities to choose from for every query, resulting in $P^Q$ possibilities overall\footnote{If no savings were given, the least-cost solution is trivially obtained by selecting the least-cost plan of every query.}.
Informally, we can formulate the following rules to constrain the search space:

\begin{enumerate}
    \item[$\mathcal{R}_1$] Plans with lower cost are preferred.
    \item[$\mathcal{R}_2$] Cost savings across any queries are taken into consideration.
    \item[$\mathcal{R}_3$] Exactly one plan is selected per query.
\end{enumerate}

\begin{example}
We consider two queries $q_1$, produced by plans $p_1$ and $p_2$ and $q_2$, produced by plans $p_3$ and $p_4$. The cost for the plans $p_1, p_2, p_3, p_4$ are $3, 13, 21$ and $1$. When plans $p_2$ and $p_3$ are both selected, they save costs of $s_{2,3} = 14$. To find the least-cost solution, we compute the cost of every admissible selection as displayed below. That is, we only list solutions that feature exactly one plan per query.
Obviously, selecting plans $p_1$ and $p_4$ is the least-cost solution. \\

\begin{centering}
\begin{tabular}{c c l}
    $q_1$ & $q_2$ & Cost \\
    \textcolor{gray}{$p_1$}, \textcolor{red}{$p_2$} & \textcolor{gray}{$p_3$}, \textcolor{red}{$p_4$} & $13 + 1 = 14$ \\
    \textcolor{gray}{$p_1$}, \textcolor{red}{$p_2$} & \textcolor{red}{$p_3$}, \textcolor{gray}{$p_4$} & $13 + 21 - 14_\text{(Savings)} = 20$ \\
    \textcolor{red}{$p_1$}, \textcolor{gray}{$p_2$} & \textcolor{gray}{$p_3$}, \textcolor{red}{$p_4$} & $3 + 1 = 4$ \\
    \textcolor{red}{$p_1$}, \textcolor{gray}{$p_2$} & \textcolor{red}{$p_3$}, \textcolor{gray}{$p_4$} & $3 + 21 = 24$ \\
\end{tabular} \\
\end{centering}
\end{example}

\subsection{Contributions of the Paper}

The MQO scales exponentially with the number of query plans it contains \cite{trummer2015dwave}. While classical computing typically approaches such problems by pruning the search space \cite{trummer2015dwave}, quantum computers can explore the entire search space simultaneously \cite{nielsen2002quantum,moll2018quantum}.

In Section \ref{sec:quantum_computing} we revise the most important foundations of quantum computing that are important for solving optimization problems such as multiple-query optimization.

In Section \ref{sec:background} we first show a classical brute-force approach for the MQO that enumerates and then evaluates every solution. We then provide an intuition for the quantum query optimization by characterizing the optimal solution in a way that can later be used for a quantum circuit, which is the quantum-equivalent to classical machine language. Finally, we show a high-level overview of our algorithm.

In Section \ref{sec:algorithm}, we first provide the necessary conceptual groundwork for the Quantum Approximate Optimization Algorithm (QAOA), on which the algorithm is based on, and establish the relationship to quantum annealing.
Next, we formulate the MQO mathematically, translate it into a quantum circuit while simultaneously introducing the classic part of our algorithm.
We conclude that section with a more refined low-level algorithm and a discussion about how the algorithm is run.

Finally, in Section \ref{sec:results}, we present experimental results of our algorithm. As the algorithm's performance relies heavily on the classical optimization part, we utilize novel optimization strategies that we analyze.
Furthermore, we compare our approach with the quantum annealing approach to the MQO and demonstrate that our approach shows more efficient usage of qubits which is an advantage for future, more powerful gate-based quantum computers. We complete our results with a complexity analysis of our approach.

In summary, our contributions are the following:
\begin{enumerate}
    \item We provide a novel hybrid classical-quantum algorithm to solve the multiple query optimization. The only previous quantum computing-based approach tackling the multiple query optimization used a quantum annealer architecture \cite{trummer2015dwave}. To the best of our knowledge, this is the first paper in the literature to address the multiple query optimization problem with a gate-based quantum computer.
    \item We conduct experiments with the algorithm with real quantum hardware and simulators, and analyze how the algorithm scales with the problem size.
    \item We compare our work with a competing quantum query optimization approach.
\end{enumerate}

\section{Quantum Computing Foundations}
\label{sec:quantum_computing}

Quantum algorithms are known for speeding up certain computations, such as integer factorization \cite{nielsen2002quantum}. In principle, these algorithms achieve the significant speedup by not having to examine every branch in the search space sequentially, as by a classical brute force approach\footnote{The conception of a quantum computer evaluating the entire search space in parallel and selecting the best solution is an incorrect oversimplification \cite{aaronson2013quantum}.}.
Instead, a quantum computer holds a large number of branches during computation, which is possible due to the fact that a quantum register can be understood as a superposition of basis states. 

Here, we must add a brief excursion. Certain types of quantum systems, e.g. electrons or polarized photons, are in quantum mechanics represented by a data structure that is equivalent to a normalized vector in a two-dimensional complex vector space (i.e. a two-dimensional complex Hilbert space, to be precise). Such vectors, denoted by a so called ket $\ket{q}$, are the superposition of two basis vectors $\ket{0}$ and $\ket{1}$:
\begin{equation}
\ket{q} = \alpha \ket{0} + \beta \ket{1} 
\end{equation}
whereby $\alpha,\beta \in \mathbb{C}$ and 
\begin{equation}
\ket{q} = | \alpha |^2 |\beta|^2 = 1. 
\end{equation}

The data structure $\ket{q}$ is called a qubit. Without delving into mathematical and physical details: qubits can be combined and then form a so called \textit{quantum register}. If one has $n$ coupled qubits, the resulting quantum register is again a normalized element of a complex Hilbert space, whereby this space has dimension $2^n$. A basis state of this Hilbert space can be understood as a bit sequence $\ket{x_0 x_1 ... x_n}$ with $x_i \in \{0,1\}$. The bit sequence of the basis state encodes a natural number $i$, therefore the basis states are often just written as $\ket{i}$ with $i \in {0,...,2^n - 1}$. A quantum register $\ket{Q}$ is then a superposition of basis states:

\begin{equation}\label{qstate}
\ket{Q} = \sum_{i=0}^{2^n} c_i \ket{i}  
\end{equation}
with 
\begin{equation}
\sum_{i=0}^{2^n} |c_i|^2 = 1  
\end{equation}

One aspect of superposition is so called \textit{quantum parallelism}: One can understand an operation on a quantum register as the simultaneous operation on many basis states. This interpretation is sensible, because quantum operations are (in a mathematical sense) linear operations in a vector space that is built up from basis states. Another aspect (not explicitly discussed in this paper but relevant for many theoretical features of quantum computing) is due to the fact that quantum registers represent entangled states. A quantum computation is ended by some sort of quantum measurement. In brief, a quantum algorithm manipulates quantum registers with the goal of amplifying the probability of measuring the basis state (bit strings) that represents the solution to the problem under investigation. 

The time development of a quantum state is described by the Schrödinger equation:
\begin{equation}\label{schroed}
i \hbar \frac{\partial}{\partial t} \ket{Q} = H \ket{Q}. 
\end{equation}
Thereby, $H$ is an operator (here a $2^n$-dimensional matrix). The matrix $H$ is called the \textit{Hamiltonian} of the system and is related to the system's energy. Mathematically, $H$ is a so - called \textit{Hermitian matrix}, which means that its eigenvalues are all real numbers. These eigenvalues are the energies of the respective eigenstate: If $\ket{x}$ is an eigenstate of $H$ and $\lambda_x$ the eigenvalue of  $\ket{x}$, then the system carries energy $\lambda_x$ if it is in state $\ket{x}$. The state with the lowest energy (the smallest $\lambda_x$) is called the ground state. 

The time evolution described by Equation \ref{schroed} is not the only process of relevance in quantum computing. There are also so called \textit{measurements}. For our purposes, it is sufficient to define a measurement as a stochastic, non-linear operation that transforms a general quantum register $\ket{Q}$ into one, randomly chosen basis state. Formally, a measurement $M$ is defined by (we use the notation of Equation \ref{qstate}):

\begin{equation}\label{measurement}
M \ket{Q} =  \ket{i} \qquad \text{with probability $|c_i|^2$}. 
\end{equation}

This means that a measurement destroys the information stored in a quantum register, at least to some extent. That is the reason, why many algorithms in quantum computing can be understood as \textit{amplification processes} for the basis state that represents the solution one is looking for (or geometrically speaking, rotate a general state towards the solution state by reducing the angle between a general state and the solution). 

Equation \ref{schroed} describes the temporal development of the system under consideration. One can show that the fact that $H$ is Hermitian leads to time evolution that preserves the length of the vector describing the state of the system. An operation that preserves the length of a vector is simply a rotation (though in high-dimensional vector space). An operator that mediates such a rotation is called \textit{unitary}. The time evolution of a quantum mechanical system is therefore a unitary process. Process steps in quantum computing consist of unitary operations; one physical challenge one faces is to implement the Hamiltonian $H$ that belongs to a desired unitary operation $U$.   

Note well that the unitary time evolution given by Equation \ref{schroed} is reversible, whereas a measurement by Equation \ref{measurement} is an irreversible process. 

Finally, we point out that even if the Hilbert space that describes a quantum register may be of high dimension, it is still a \textit{finite dimensional space}. That means that the mathematics one needs for quantum computing is (comparably) simple; linear algebra is sufficient. The subtleties of linear functional analysis (colloquially linear algebra in infinite dimensional spaces) are no barrier for understanding most algorithms in quantum computing. 

\begin{figure}
\centering
\includegraphics[width=0.75\linewidth]{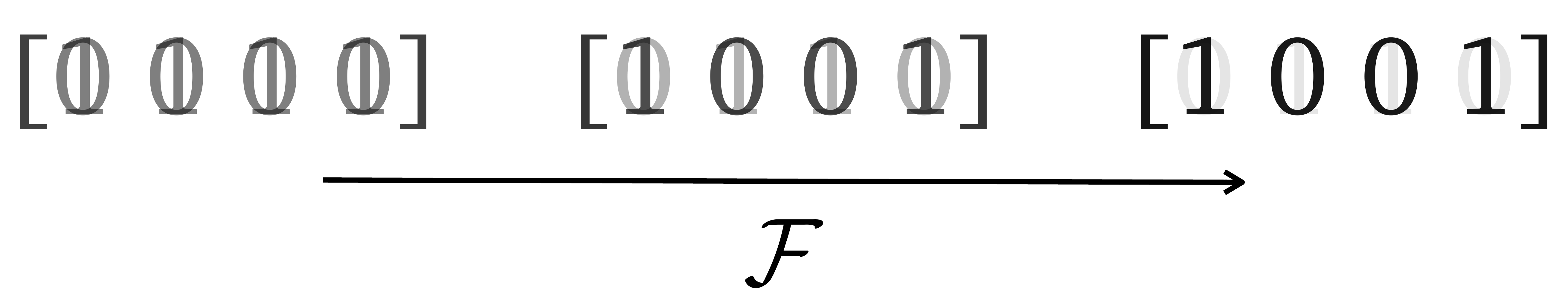}
\caption[The adiabatic evolution]{$\mathcal{F}$ gradually augments the probability that a desired solution is measured.}
\label{fig:quantum_evolution}
\end{figure}

\section{Quantum Query Optimization: High Level Overview}\label{sec:background}

Let us now discuss how to solve the problem of multiple-query optimization using the theory revised in the previous section.

We define a black-box function $\mathcal{F}$ that encodes solutions in the sense that it alters a superposition of basis states in such a manner that those representing solutions to a given problem are amplified. We call $\mathcal{F}$ a \textit{problem encoding function}. An intuitive interpretation of the process to find an optimal solution is rendered in Figure \ref{fig:quantum_evolution}.

There is a common misunderstanding about $\mathcal{F}$. The existence of $\mathcal{F}$ does not imply that one "knows the answer to a problem in advance". If compared to a classical computer and a classical problem such as adding to numbers, a classical analogue for a problem encoding function is a hard-wired adder. The adder "knows" the answer to an addition in the sense that it implements the logical processes necessary for an addition. By no means, it contains the answer to all possible additions in the sense of a table.

To encode solutions to the MQO, we choose bit strings in the form $[p_1, p_2, \ldots, p_{PQ}]$, where a 1 at the $i$-th position denotes that query plan $p_i$ has been selected for the solution, and a 0 otherwise. For instance, selection of plans $p_1$ and $p_4$ is then represented as $[1, 0, 0, 1]$.

\subsection{Classical Brute Force Query Optimization}

A classical brute-force approach to find a least-cost solution for a MQO is characterized as follows:
\begin{enumerate}
    \item Get bit strings that encode admissible solutions.
    \item Evaluate the cost of each solution.
    \item Return the solution with minimum cost.
\end{enumerate}
A pseudocode-implementation of such a program is shown in Listing \ref{lst:mqo_pseudo}.

{\small
\begin{lstlisting}[caption={Program structure to find the optimal solution}, label={lst:mqo_pseudo}]
# 1.
solutions = get_admissible_solutions()
# 2.
for s in solutions:
  solution_cost[i] = cost(s) - savings(s)
# 3.
min_idx = argmin(solution_cost)
return solutions[min_idx]
\end{lstlisting}
}

\subsection{Intuition for Quantum Query Optimization}
In this section, we provide an intuition for the hybrid quantum algorithm. The details will be discussed in Section \ref{sec:algorithm}. Therefore, we neglect theoretical rigor in favor of a better introduction to the algorithm.

We choose the number of qubits in the quantum register equal to the number of plans of the MQO, as we want the quantum algorithm to operate with bit strings that encode solutions\footnote{In this section we are using the terms "bit string" and "solution" interchangeably, as bit strings encode solutions.}. Initially, the quantum register can be prepared to contain a uniform superposition of  all $2^n$ possible \textit{basis states} with equal probability of being measured. 

\begin{figure}
    \centering
    \includegraphics[width=0.9\linewidth]{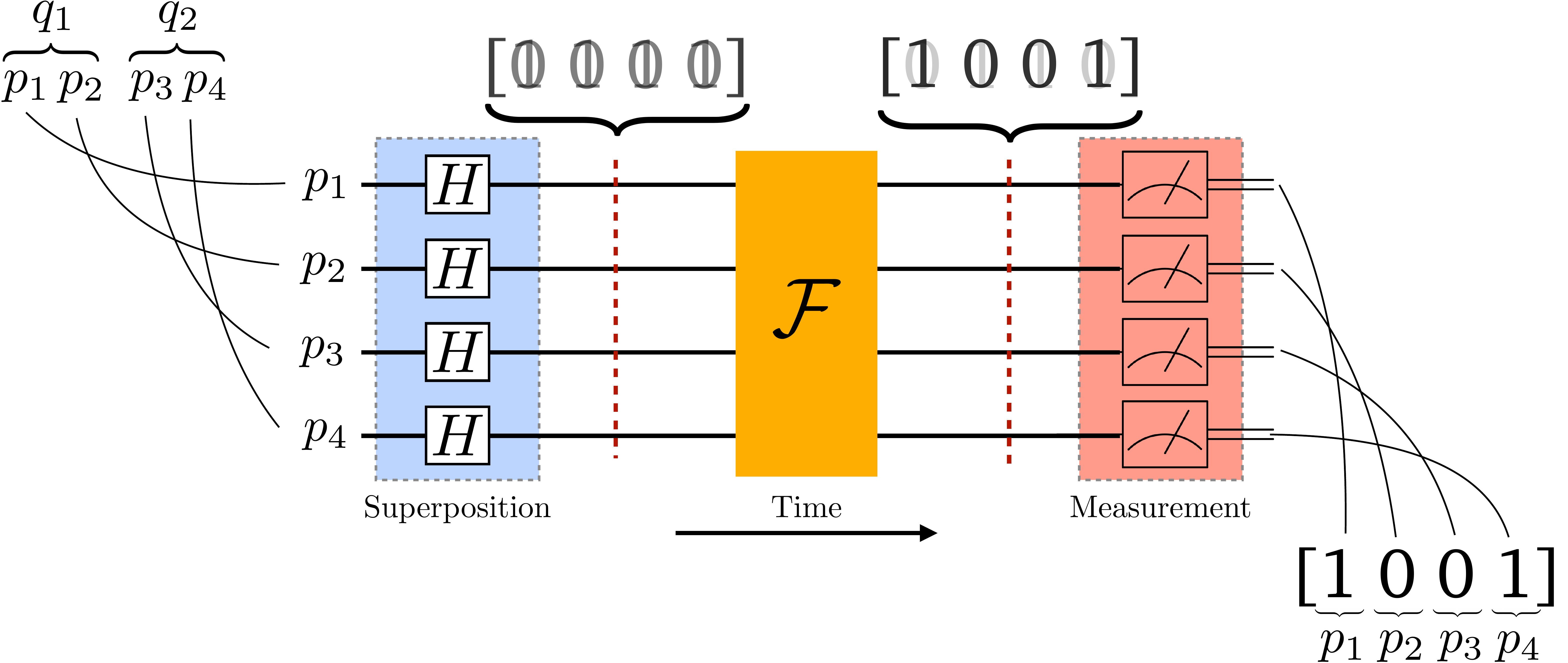}
    \caption[Overview of the high level quantum query optimization algorithm]{The structure of the algorithm. First, qubits are put into uniform superposition (blue), $\mathcal{F}$ is applied, driving the superposition towards the optimal solution, and finally, the basis state amplitudes are measured (red).}
    \label{fig:algo_structure}
\end{figure}

A classical computer, even when employing heuristics, must evaluate individual bit strings from the search space. The operator $\mathcal{F}$ acts on the entire quantum register and thus operates on $2^n$ states simultaneously, as rendered in Figure \ref{fig:algo_structure}.
According to its implementation, $\mathcal{F}$ augments the amplitude of certain states, which are then more likely to be measured. The effect is insofar somewhat similar to that of a genetic algorithm: solutions with desired properties are promoted, where others are disregarded. However, the process is completely deterministic and, as mentioned, from a mathematical perspective nothing else than a rotation in a high dimensional vector space. 

Internally, $\mathcal{F}$ is constructed from quantum gates and encodes a particular problem. In case of the MQO, we have already defined the rules that describe desirable solutions (see Section \ref{sec:introduction} after Example 1).

The encoding rewards desired properties e.g. savings, and penalizes undesired ones, e.g. high plan costs. Non-admissible solutions that encode no or multiple plans for a query are non-negotiable and therefore receive a dominant penalty. By combining these rewards and penalties in $\mathcal{F}$, all necessary information to promote the optimal solution is encoded. The encoding is formally discussed in Section \ref{sec:algorithm}.

With $\mathcal{F}$ being at the core of the quantum algorithm, it requires two further components: the uniform superposition on which $\mathcal{F}$ operates and the measuring step, which delivers, at least with a sufficiently high probability, the basis state representing the solution.

Our quantum algorithm can thus be summarized as succession of four steps: (1) Encode the problem. (2) Put every qubit in the quantum register into a superposition to acquire an uniform superposition. (3) Apply $\mathcal{F}$ on the quantum register. (4) Measure the state of the qubits. Figure \ref{fig:algo_structure} gives an overview of the algorithm's circuit.

A high-level quantum multiple query optimization algorithm can be formulated as in Algorithm \ref{algo:highlevel}. 

\begin{algorithm}
\SetAlgoLined
\SetKwInOut{Input}{Input}\SetKwInOut{Output}{Output}
\Input{Plan cost $C$, Savings $S$, amount of queries $Q$}
\Output{Least-cost solution}
(1) Encode problem into $\mathcal{F}$ from $C$, $S$ and $Q$\;
(2) Create uniform superposition on quantum register\;
(3) Apply $\mathcal{F}$ on quantum register\;
(4) Measure quantum register\;
\Return{Least-cost solution}\;
 \caption[High level quantum query optimization algorithm]{High level quantum query optimization algorithm.}
 \label{algo:highlevel}
\end{algorithm}

This chapter omitted details such as the number of times $\mathcal{F}$ has to be applied, or the implementation details of $\mathcal{F}$. Both of these topics are refined in the next section.

\section{Hybrid Classical-Quantum Algorithm} \label{sec:algorithm}
The intuition of the previous section outlines the basic workings of the algorithm for multiple query optimization. In this section we develop this intuition into a more rigid framework considering the practical limitations of current quantum hardware. For this, we first introduce the context of hybrid classical-quantum algorithms.

\subsection{Toward a Quantum Approximate Optimization Algorithm}
The class of \textit{hybrid classical-quantum algorithms} utilizes both classical and quantum computers. These types of algorithms are considered promising since they can better handle erroneous, small-scale quantum devices \cite{moll2018quantum} than "pure" quantum algorithms.
QAOA has been suggested as one of the major hybrid classical-quantum algorithms \cite{farhi2014quantum}, next to the \textit{Variational Quantum Eigensolver} (VQE) \cite{peruzzo2014variational}. The former is often used for generic optimization problems while the latter is often used in context of molecule simulations. In this work, we use the QAOA.

The quantum mechanical foundation that enables a function like $\mathcal{F}$ to "select" optimal solutions is the \textit{adiabatic principle} \cite{born1928beweis, farhi2000quantum}. It (roughly) states that a quantum mechanical system, evolved by a gradually applied time dependent operator, remains in its instantaneous eigenstate \cite{farhi2000quantum}. 

This means that if the system starts in a known ground state and the problem-encoding function $\mathcal{F}$ is applied gradually, it ends in the state that generates the least cost for our problem \cite{homeister2008quantum}. This evolution is informally comparable to learning a neural network. Although the form of the neural network is unknown and must first be learned, the properties of neural network are known, namely that it maps an input to an output with minimal loss. Similarly, the evolution's goal state can be described by a problem-encoding function as $\mathcal{F}$.

\textit{Quantum annealers} are a class of quantum computers that natively facilitate this evolution \cite{trummer2015dwave, dwave}. Quantum annealers typically comprise more qubits than \textit{gate-based} quantum computers \cite{ibmq-specs,dwave}. While quantum annealers only facilitate computations of a certain class of problems, gate-based quantum computers are more general, as they can implement any quantum algorithm. In this work, we focus solely on gate-based quantum computers. 

\begin{figure}
    \centering
    \includegraphics[width=0.8\linewidth]{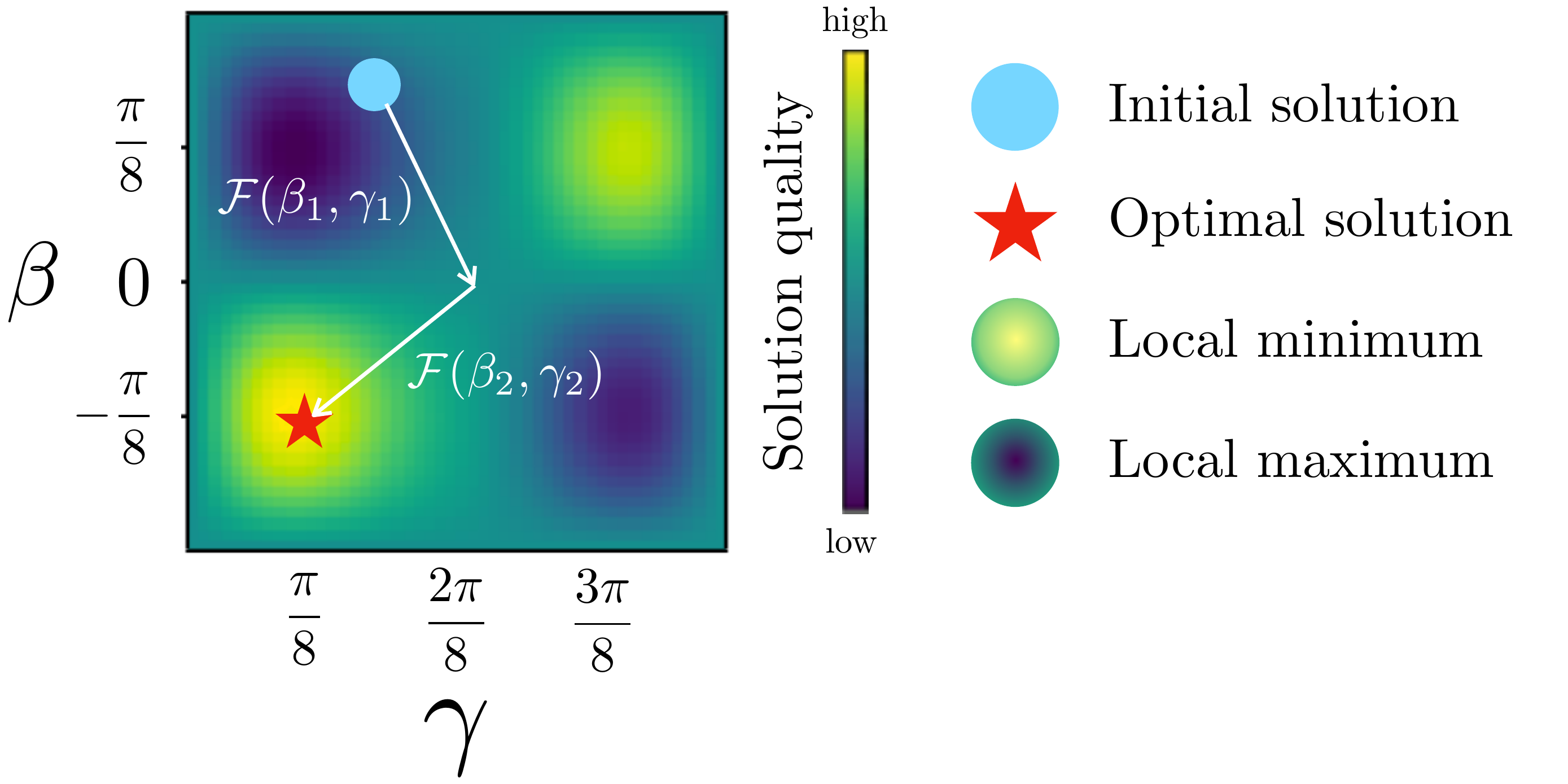}
    \caption[Solution space with parameters]{The solution space is represented in dependence of parameters $\beta, \gamma$ used for $\mathcal{F}$. The goal is to find a set of parameters that guides $\mathcal{F}$ to a global minimum, where the desired solution lies. $\mathcal{F}$ is decomposed into $p=2$ steps and the chosen parameters lead it to the optimal solution.}
    \label{fig:evolution_params}
\end{figure}

To compute the evolution on gate-based quantum computers, the evolution process is decomposed into discrete units \cite{farhi2000quantum}.
In case of our problem-encoding function $\mathcal{F}$, the decomposition is realized by applying $\mathcal{F}$ step-wise with the two parameters $\beta$ and $\gamma$ describing the "step size". $\mathcal{F}$ can be expressed as follows:

\begin{equation}
    \mathcal{F} \xrightarrow[]{\text{Discretization}} \mathcal{F}(\beta_1, \gamma_1) + \mathcal{F}(\beta_2, \gamma_2) + \ldots + \mathcal{F}(\beta_p, \gamma_p)
\end{equation}

where $p$ denotes the amount of decomposable units of $\mathcal{F}$. The parameters $\beta$ and $\gamma$ influence significantly how well the evolution into the goal state described by $\mathcal{F}$ succeeds (see Figure \ref{fig:evolution_params}). Using the neural network analogy, the parameters can be seen as a learning rate. Chosen too steep, it "overshoots" the optima \cite{buduma2017fundamentals}.

The decomposition of $\mathcal{F}$ typically leads to many gates and thus, deep quantum circuits.
As the error rate grows exponentially with the depth of the circuit and robust error correction is currently out of reach, deep quantum circuits are infeasible.

The QAOA tackles this problem by choosing the parameters $\beta, \gamma$ such that $\mathcal{F}$ is most effectively applied and the evolution succeeds, approximating the optimal cost value. 
Thereby, it exploits the strengths of quantum computers (i.e. fast evaluation and selection of promising solutions in high dimensional space), while delegating tasks in which they perform poorly to classical computers (i.e. evaluating solutions to improve $\beta$ and $\gamma$).

\begin{figure}[h]
    \centering
    \begin{subfigure}[b]{0.4\textwidth}
        \centering
        \includegraphics[width=\textwidth]{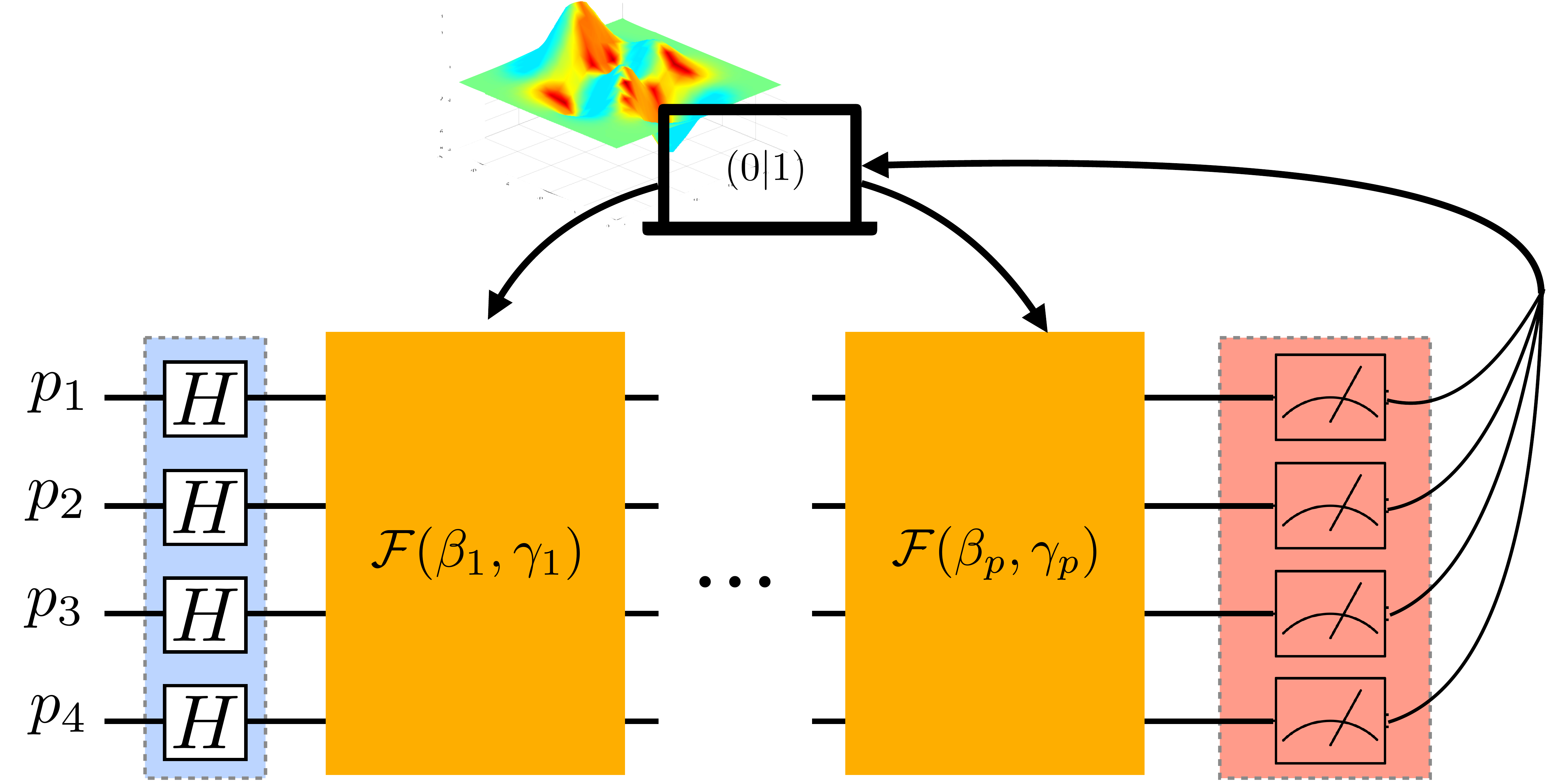}
        \caption{}
    \end{subfigure}
    \begin{subfigure}[b]{0.3\textwidth}
        \centering
        \includegraphics[width=\textwidth]{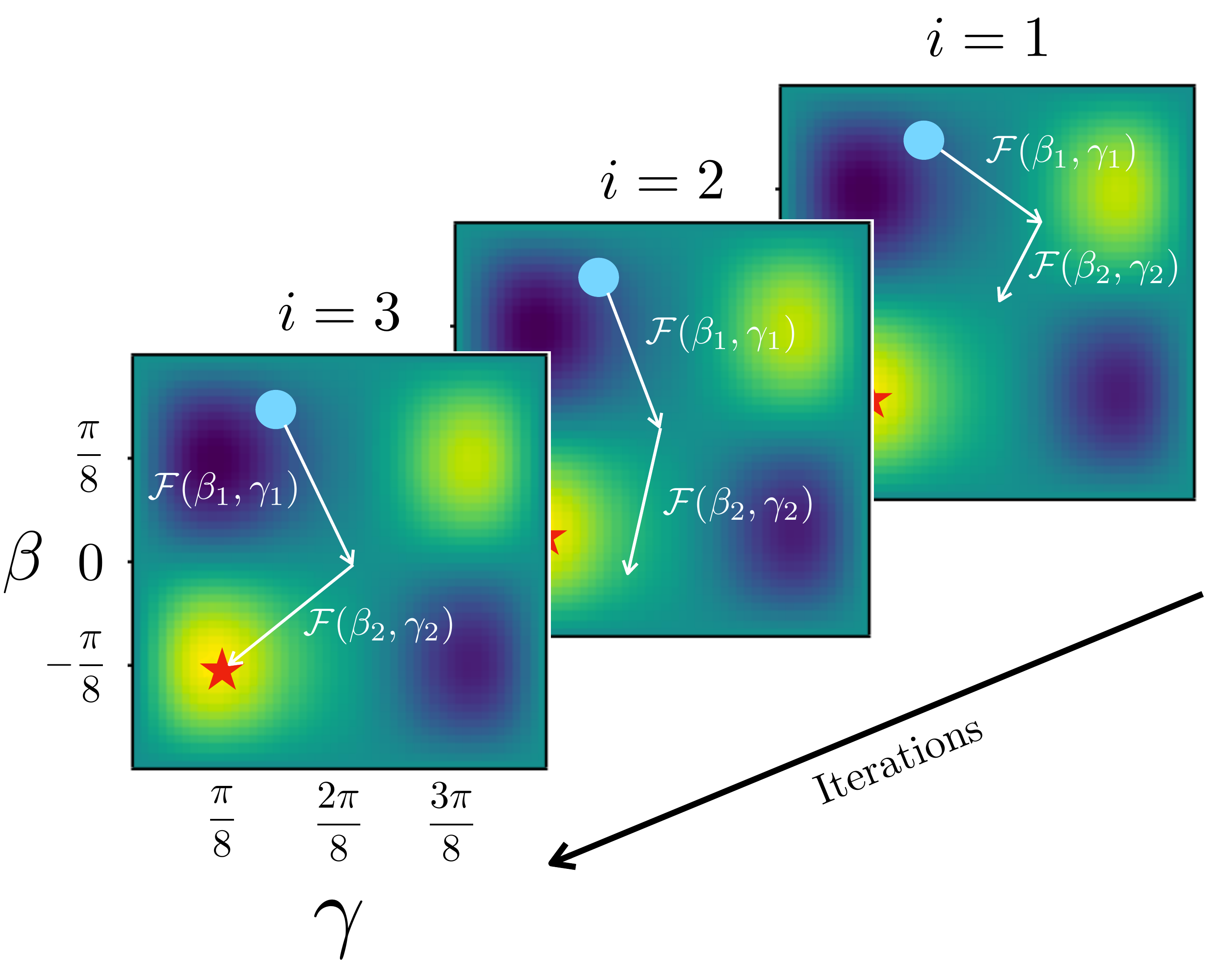}
        \caption{}
    \end{subfigure}
    \caption[Structure and procedure of QAOA]{(a) The QAOA's quantum circuit prepares the uniform superposition by applying $H$ gates, followed by parametrized application of $\mathcal{F}(\beta, \gamma)$ and concludes by measuring the quantum register's state. Each query plan $p_1, \ldots, p_4$ is encoded by a qubit. After each iterator of the algorithm, the parameters are optimized by a classical computer and fed back to the quantum circuit. (b) Parameters are improved over multiple iterations. Each iteration allows $\mathcal{F}$ to better reach the optimal solution.}
    \label{fig:QAOA_overview}
\end{figure}

A summary of the QAOA algorithm is given in Figure \ref{fig:QAOA_overview} (a). It consists of a quantum part that applies a problem-encoding function $\mathcal{F}$ step-wise on a quantum register, which is first brought into uniform superposition by $H$-gates (blue, so-called Hadamard - gates). Each qubit encodes a single plan, thus the quantum register can represent any solution to the MQO.
Because $\mathcal{F}$ is applied step-wise, multiple parametrized functions $\mathcal{F}(\beta, \gamma)$ appear on the quantum circuit, each time with a different set of parameters. The number of appearances is denoted as $p$.
Finally, the quantum circuit is measured (red) and the resulting solutions are fed to the classical part. Based on the quality of the solution\footnote{It is computationally inexpensive to evaluate a solution, see Section \ref{sec:background}.}, the classical computer calculates a new set of parameters $(\beta_1, \gamma_1), \ldots, (\beta_p, \gamma_p)$ which are expected to evolve $\mathcal{F}$ more effectively, thus producing better solutions.
As a classical optimization technique, gradient descent may be used \cite{zhou2020quantum, moll2018quantum}.
This closed loop is repeated $I$ times until convergence, or a threshold is reached (see Fig. \ref{fig:QAOA_overview} (b)).
The optimization of $\beta$ and $\gamma$ is crucial to the success of the algorithm and is discussed at the end of this section.

\subsection{Revealing $\mathcal{F}$'s Problem Encoding}
Until now, $\mathcal{F}$ has been treated as black-box function, encoding the problem with qualitative constraints previously defined in the form of rewards and penalties. Quantitatively, the constraints can be formulated in terms of a classical \textit{cost function} to be minimized, similar to \cite{trummer2015dwave}. To distinguish this cost function from the intuitive problem-encoding function $\mathcal{F}$, we subsequently call the classical cost function $\mathcal{F}_C$.
The variables of the cost functions are equal to the solution-encoding bit strings (see Section \ref{sec:background}) in the form $[p_1, p_2, p_3, p_4]$ e.g. $[0, 1, 1, 0]$.

For the MQO, the cost function ${F}_C$ encompasses three sums for the constraints and is defined as follows according to \cite{trummer2015dwave}. The sums are directly derived from the rules $\mathcal{R}_1$ to $\mathcal{R}_3$ defined in Section \ref{sec:introduction}.

\begin{equation}
    {F}_C = {S}_1 + {S}_2 + {S}_3
\end{equation}

$S_1$, $S_2$ and $S_3$ are defined as follows:

\begin{equation}
    {S}_1 = \sum_{i = 1}^P p_i \cdot (c_i - w_{\min})
\end{equation}

\begin{equation}
    {S}_2 = -\sum_{s \in S} s_{i,j} \cdot p_i \cdot p_j
\end{equation}

\begin{equation}
    {S}_3 = \sum_{i = 1}^Q \sum_{p_k, p_j \in q_i} w_{\max} \cdot p_k \cdot p_j
\end{equation}

\begin{equation}
    w_{\min} = \max(C) + \epsilon
\end{equation}

\begin{equation}
    w_{\max} = w_{\min} + \sum_{s \in S} s 
\end{equation}


We will now analyze the three sums $S_1$, $S_2$ and $S_3$ in more detail.

$S_1$ calculates the \textit{costs} $c_i$ of all query plans $p_i$. The term $w_{\min}$ gives an incentive to select at least one plan. Moreover, the constant $\epsilon$ ensures that $w_{\min}$ is slightly larger than the plan with the maximum cost and guarantees that at least one plan is selected for every query. 

${S}_2$ calculates the total \textit{savings} $s_{i,j}$ of shared resources between query plans $p_i$ and $p_j$, i.e. ${S}_2$ rewards savings if the plan pairs $p_i, p_j$ are selected, for which savings $s_{i,j}$ are applicable.

${S}_3$ penalizes solutions that select \textit{more than one plan per query}. As this is a non-negotiable constraint, it is multiplied by $w_{\max}$ to dominate any bonus.


\begin{figure*}
    \centering
    \includegraphics[width=\textwidth]{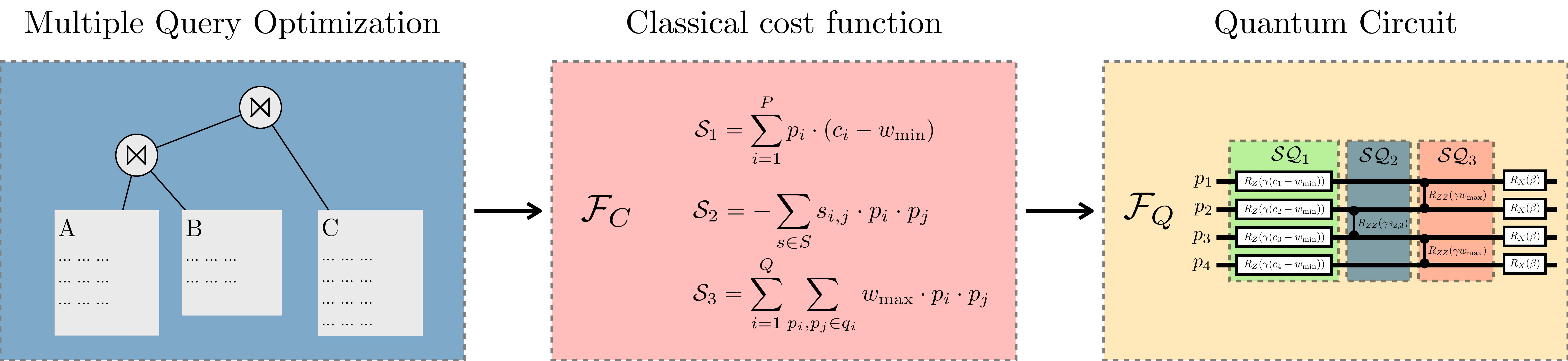}
    \caption[Overview of the problem encoding]{Overview of the problem encoding. First, the MQO is formulated into a quadratic classical cost function (QUBO) $\mathcal{F}_C$. Due to the similarity with quantum gates, this function is finally implemented as a quantum circuit $\mathcal{F}_Q$ with rotational gates.}
    \label{fig:encoding_overview}
\end{figure*}

\begin{example}
To illustrate the cost function, we apply it on the example MQO. Recall that the costs for the plans $p_1, p_2, p_3, p_4$ are $3, 13, 21$ and $1$ - as shown previously in Example 2. When the plans $p_2$ and $p_3$ are selected, they have a cost saving of $s_{2,3} = 14$. For this example we set $\epsilon$ to 1. For all solutions, we calculate $w_{\min} = \max(3, 13, 21, 1) + 1 = 22$ and $w_{\max} = w_{\min} + 14 = 36$.
$[p_1, p_2, p_3, p_4] = [1, 0, 1, 1]$ is a non-admissible solution, as it selects plans $p_1, p_3$ and $p_4$, where $p_3$ and $p_4$ produce the same query $q_2$.

For the solution, we calculate $\mathcal{S}_1 = (3-22) + (21-22) + (1-22) = -41$. Then, $S_2$ is calculated easily as there are no savings between the selected plans hence $\mathcal{S}_2 = 0$. Finally, $\mathcal{S}_3 = w_{\max} \cdot p_3 \cdot p_4 = 36$, which represents the penalty due to two selected plans for the same query. The complete sum is given by $\mathcal{S}_1 + \mathcal{S}_2 + \mathcal{S}_3 = -41 + 0 + 36 = -5$.

Now, we calculate the cost for the valid and optimal solution $[p_1, p_2, p_3, p_4] = [1, 0, 0, 1]$. $\mathcal{S}_1$ = (3-22) + (1-22) = -40. Because no savings apply between plans $p_1$ and $p_4$ we set $\mathcal{S}_2 = 0$. $\mathcal{S}_3 = 0$ as well, because $w_{max} \cdot p_1 \cdot p_2 = 36 \cdot 1 \cdot 0 = 0$ and $w_{max} \cdot p_3 \cdot p_4 = 36 \cdot 0 \cdot 1 = 0$, therefore there is no pair of $p_i, p_j$ for the same query whose product is 1. Thus, the sum of $\mathcal{S}_1 + \mathcal{S}_2 + \mathcal{S}_3 = -40 + 0 + 0 = -40$.

The second solution with $-40$ yields a lower value than the first solution with $-5$, which is desired for minimization of the cost function. Consequently the query plan $[1, 0, 0, 1]$ is preferred over the query plan $[1, 0, 1, 1]$.
\end{example}

As the sums $\mathcal{S}_2$ and $\mathcal{S}_3$ are quadratic terms (i.e. they contain the product of the two variables $p_i, p_j$), this cost function is classified as a \textit{quadratic cost function}. In the literature, this class of problem is also known as \textit{Quadratic Unconstrained Binary Optimization} (QUBO) \cite{trummer2015dwave,gilliam2021grover}.
Problems in the QUBO formulation have a similar form to problems solvable by Quantum Computers\footnote{More precisely, these are \textit{Ising models}, described by \textit{Hamiltonian} matrices\cite{lucas2014ising}.} \cite{Qiskit,glover2019tutorial}.
Here, we omit the quantum mechanical intricacies, but refer to \cite{farhi2014quantum, moll2018quantum} for a more detailed explanation of this relationship.

\subsection{Implementing $\mathcal{F}_C$ as Quantum Circuit}\label{ssec:implementing}
On a practical level, the relation between the QUBO form of $\mathcal{F}_C$ and quantum computing is due to quantum gates that typically act on one or two qubits and QUBO problems that feature one or two variables in every term.
Hence, linear variables of the classical cost function are translated into two standard gates in quantum computing, a quantum rotation gate $R_{Z}(\theta)$ with the parameter $\theta$ being proportional to the scalar (such as cost) the variable is multiplied with.
Quadratic variables are implemented using $R_{ZZ}(\theta)$-gates (another standard type of gate) acting on two qubits, representing the two variables.
We subsequently call the quantum-gate implementation of the classical cost function $\mathcal{F}_Q$, or quantum cost function.

\begin{example}
To illustrate the implementation consider the example MQO. First, the linear (quantum) sum $\mathcal{S}_1$ is implemented:

\begin{equation}
\mathcal{SQ}_1 = \sum_{i = 1}^N R_{Z}(\gamma (c_i - w_{\min})) \ket{p_i}
\end{equation}

acting on the qubits $\ket{p_i}$ that encode the plans.
Note that the factor $\gamma$ controlling the step size for gate-based quantum computer is encompassed as well. Here, this sum produces one $R_Z$-gate for every plan i.e. qubit. Assuming $\gamma = 0.5$, we obtain for the first plan that has cost of $3$ and with $w_{\min}$ that is $22$ the gate $R_{Z}(0.5 \cdot (3 - 22)) \ket{p_1}$, where $\ket{p_1}$ is the qubit the gate is applied to. The plans $p_2$ to $p_4$ are created in the same manner.

Quadratic sums are implemented analogously, as for instance with $\mathcal{SQ}_2$, which is internally quadratic:

\begin{equation}
   \mathcal{SQ}_2 = -\sum_{s \in S} R_{ZZ}(\gamma s_{i,j}) \ket{p_i, p_j}
\end{equation}

This gives us a gate for every saving, which is pairwise between plans. As there is only a single saving between plans $p_2$ and $p_3$ with the value $14$, the quantum gate is $R_{ZZ}(0.5 \cdot (-14)) \ket{p_2, p_3}$. Note that the negative sign of the sum has been taken in front of the saving's cost as gate parameter.

Lastly, $\mathcal{SQ}_3$ is implemented as follows:

\begin{equation}
    \mathcal{SQ}_3 = \sum_{i = 1}^Q \sum_{p_i, p_j \in q_i} R_{ZZ}(\gamma w_{\max}) \ket{p_i, p_j}
\end{equation}

This sum typically contributes the majority of gates as all plans within a query must be pairwisly connected. Recall that $w_{\max}$ equals to $36$. For query $q_1$ this yields the gate $R_{ZZ}(0.5 \cdot 36) \ket{p_1, p_2}$ and for query $q_2$ the gate $R_{ZZ}(0.5 \cdot 36) \ket{p_3, p_4}$.

A general analysis of the amount of quantum gates depending on the problem size is given at the end of Section \ref{sec:results}.
\end{example}

\begin{figure}
    \centering
    \includegraphics[width=0.9\linewidth]{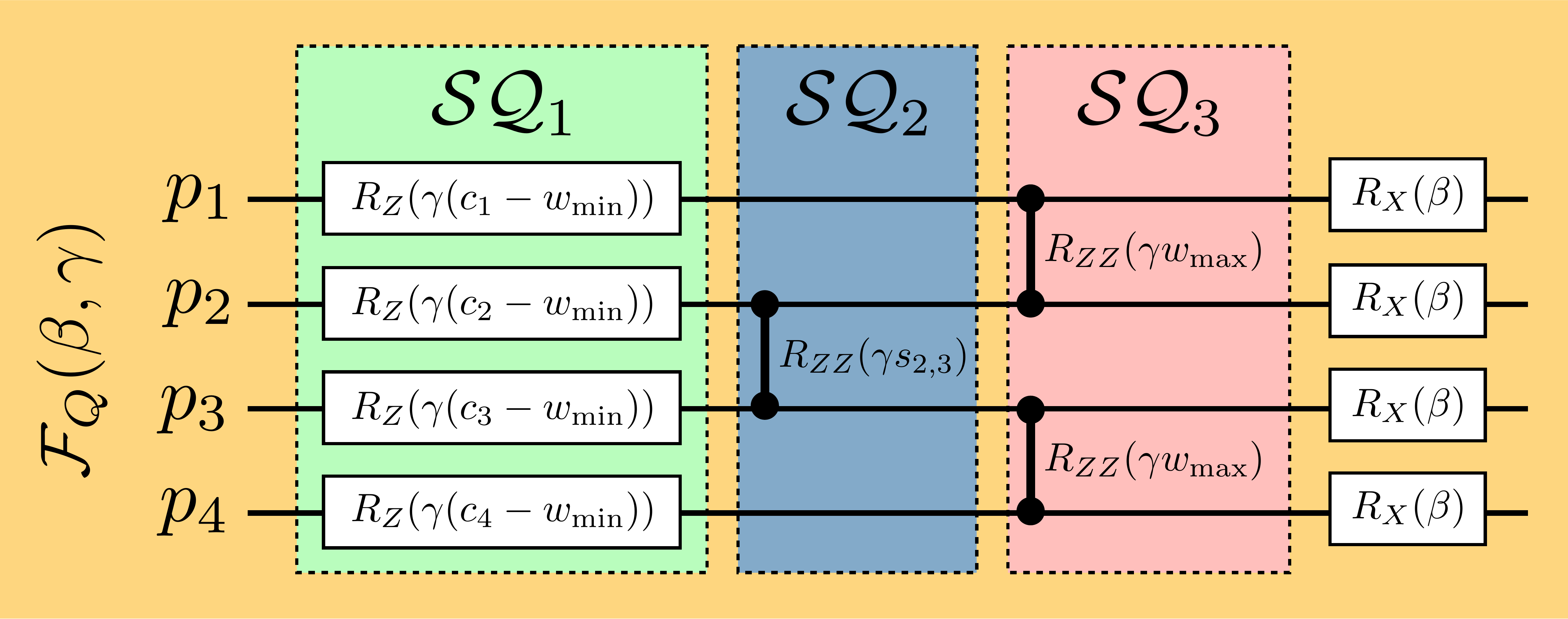}
    \caption[The final quantum circuit encoding]{The final quantum circuit encoding for $\mathcal{F}_Q$. The linear sum $\mathcal{SQ}_1$ acts with single-qubit gates that penalize (high) costs. $\mathcal{SQ}_2$ is quadratic and rewards savings on qubits $p_2, p_3$ as they apply for the corresponding plans. Finally, $\mathcal{SQ}_3$ penalizes the selection of more than one plan per query, and therefore acts on each pair of plans for a query.}
    \label{fig:f_gates}
\end{figure}

This notation is arguably cumbersome, and the encoding is better readable in the form of a quantum circuit, depicted in Figure \ref{fig:f_gates}. In difference to the classical cost function, an additional array of gates is necessary due to the adiabatic principle.
More precisely, for each qubit encoding a plan $\ket{p_1}, \ldots, \ket{p_P}$, a single $R_X(\beta)$-gate (another standard gate) is appended to $\mathcal{F}_Q$.
The exact reasons for this requirement are beyond the scope of this work and are treated in \cite{farhi2014quantum, zhou2020quantum}.
Intuitively, the $R_X(\beta)$-gates can be described as "driver" gates for the quantum annealing evolution\footnote{Therefore, the requirement is known as \textit{driver Hamiltonian}.}.

\subsection{Running the Algorithm}
The hybrid classical-quantum algorithm has two degrees of freedom. First, the number $p$ of steps $\mathcal{F}_Q$ is decomposed into and secondly, the number $I$ of times the circuit is run and parameters $\beta$ and $\gamma$ are optimized. We subsequently refer to $p$ as \textit{depth}.
Currently, QAOA is in general not well understood for depths $p>1$ \cite{zhou2020quantum}. An exception to this is the recent contribution by \textit{Arute, Frank et al.}, examining circuits with $p=3$ on Google's private Sycamore quantum device \cite{harrigan2021quantum}.
Although higher depth theoretically leads to better results, as the quantum annealing evolves "smoother" (see Fig. \ref{fig:evolution_params}), more gates contribute to more error \cite{farhi2014quantum}.
Hence, the depth $p$ should be chosen carefully depending on factors such as problem size or error rate of the quantum device.
As each additional layer of $\mathcal{F}_Q$ contributes another pair of parameters $\beta$ and $\gamma$, the parameter space to be classically optimized quickly increases, as already noted by the original authors \cite{farhi2014quantum}.
To mitigate this problem, \textit{Zhou, Wang et al.} provide helpful heuristics and strategies, which they prove by extensive experimentation on graph problems \cite{zhou2020quantum}.
We examine their contributions and their applicability to our algorithm in the next section.

The low-level hybrid algorithm for the MQO can now be summarized as follows (see Algorithm \ref{algo:lowlevel}).
In contrast to the high-level algorithm (see Algorithm \ref{algo:highlevel}), the MQO is first encoded as a classical cost function $\mathcal{F}_C$, which is then translated into a quantum cost function $\mathcal{F}_Q$ with the additional "driver" gates.
As $\mathcal{F}_Q$ is applied in $p$ steps, parameters $\beta, \gamma$ are used for the application of $\mathcal{F}_Q$ inside the quantum circuit.
After $\mathcal{F}_Q(\beta, \gamma)$ has been applied on the quantum circuit, the measurement represents a solution to the MQO problem, which we denote by $z$. The cost of $z$ can be classically evaluated as shown in Example 4.1.

The parameters are first randomly initiated, and then classically optimized by a classical computer after each measurement. The goal is to find parameters $\beta^*, \gamma^*$ that allow $\mathcal{F}_Q$ to produce the bit sequence of the least-cost solution $z^*$ on the quantum register. This goal is reached exactly when $\mathcal{F}_Q$ is applied "most effectively", as introduced at the beginning of this section.
For that, we minimize the quantum cost function $\mathcal{F}_Q$ equivalently as to minimize $z$

\begin{equation}
    \beta^*, \gamma^* = \arg \min_{\beta, \gamma} \mathcal{F}_Q(\beta, \gamma)
\end{equation}

As the algorithm is approximate, the optimization loop is typically executed until a threshold is reached instead of finding the exact solution. Often, an approximation ratio $r$ is used as figure of merit for $\mathcal{F}_Q$'s performance \cite{farhi2014quantum}. $r$ is defined as follows:

\begin{equation}
    r = \frac{\mathcal{F}_Q(\beta, \gamma)}{z_{\min}}
\end{equation}
where $z_{\min}$ denotes a lower bound for the cost of the MQO problem.

\begin{algorithm}[ht]
\SetAlgoLined
\SetKwInOut{Input}{Input}\SetKwInOut{Output}{Output}
\Input{Plan cost $C$, savings $S$, amount of queries $Q$, depth $p$}
\Output{Least-cost solution $z^*$}

 Formulate the MQO as $\mathcal{F}_C$ from $C$, $S$ and $Q$\;
 Translate $\mathcal{F}_C$ into $\mathcal{F}_Q$\;
 Define initial parameters $\beta, \gamma$\;
\While{$r >$ threshold}{
  \HiLi Put quantum register into uniform \\
  \HiLi $\:$ superposition\;
  \For{$i\leftarrow 1$ \KwTo $p$}{
  \HiLi Construct $\mathcal{F}_Q(\beta_i, \gamma_i)$ on quantum \\
  \HiLi $\:$ register\;
  }
  \HiLi $z \leftarrow$ Measure quantum register\;
  Optimize $\beta, \gamma$, e.g. with gradient descend\;
 }
 \Return{Least-cost solution $z^*$}\;
 \caption[Low level quantum query optimization algorithm]{Low level quantum query optimization algorithm. Quantum parts are highlighted in blue.}
 \label{algo:lowlevel}
\end{algorithm}

To summarize, the algorithm proceeds as follows:
\begin{enumerate}
    \item Formulate the MQO into a quantum cost function $\mathcal{F}_Q$ using a classical computer.
    \item Initialize parameters $\beta, \gamma$.
    \item Create superposition on quantum register, apply $\mathcal{F}_Q(\beta, \gamma)$ in $p$ steps on the quantum register.
    \item Measure the quantum register.
    \item Classically optimize $\beta, \gamma$ with the goal of minimizing the cost of the results and repeat from (3) until threshold is reached.
\end{enumerate}

\begin{figure}
    \centering
    \includegraphics[width=\linewidth]{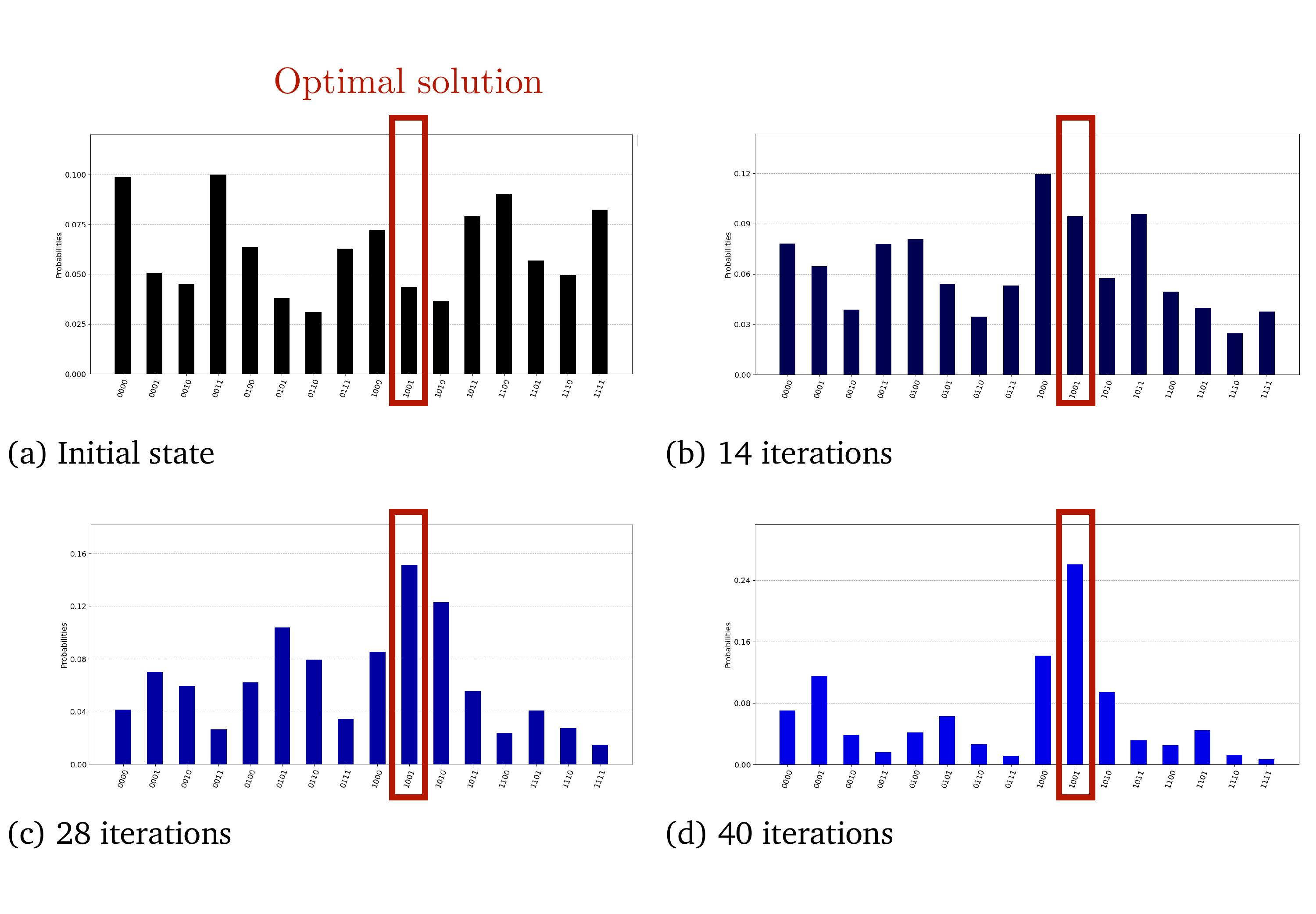}
    \caption[Development of the probability distribution of the algorithm]{The probability distribution of the measured output of the QAOA circuit of our example problem. The classical optimizer tries to find good parameters for $\gamma$ and $\beta$ to amplify the probability of measuring the solution state that produces the least costs. The probability of the correct state that minimizes the costs to the problem is clearly enhanced in (d) and therefore identified as the solution to the problem.}
    \label{fig:QAOA_evolution}
\end{figure}

Figure \ref{fig:QAOA_evolution} shows the probability distribution of the output of the circuit after 1000 measurements for selected iterations of the classical optimizer. Initially the probabilities to measure any state are distributed evenly as shown in Figure \ref{fig:QAOA_evolution} (a). After 14 iterations the most likely state, $[1, 0, 0, 0]$ would represent a non-valid solution as shown in Figure \ref{fig:QAOA_evolution} (b). Because the classical optimizer was able to avoid any local optima in this example, it minimizes the cost function even further and eventually approaches the right solution as most likely state to be measured - see Figure \ref{fig:QAOA_evolution} (c), (d). 

\section{Experiments and Results} \label{sec:results}

The primary objective of our experiments is to determine how the \textit{hybrid classical-quantum algorithm QAOA scales on current quantum computers}, i.e. how well we can solve query optimization problems. Therefore, we compare the performance of various problem sizes, i.e. various number of queries and various numbers of query plans, with different numbers of repetitions of $\mathcal{F_Q}$ - the \textit{quantum part} of the query optimization algorithm shown in Algorithm 2. $\mathcal{F_Q}$ contains the QUBO-problem to solve and adds its characteristic structure and values to the quantum circuit. 

The algorithm's "division of labor" can be summarized as follows. The parametrized \textit{quantum part} explores the search space. By exploiting quantum properties such as superposition and entanglement, the exploration is typically much faster than by classical approaches.
The \textit{classical part} is concerned with finding parameters $\gamma$ and $\beta$ that, fed to the quantum part, direct the exploration into promising regions of the search space.
With growing problem search space, the parameter search space grows as well \cite{farhi2014quantum, zhou2020quantum}, making the classical optimization challenging.

Therefore, we first analyze the classical part of the QAOA on the Qiskit quantum simulator \cite{Qiskit}. In particular, we analyze what makes good $\gamma$ and $\beta$. Afterwards, we investigate the performance of different classical optimizers. The FOURIER strategy \cite{zhou2020quantum} is a recent suggestion to improve optimization in the high-dimensional parameter space, by using a heuristic.

Finally, we evaluate the quantum part of our hybrid classical-quantum algorithm where we  run experiments on two different instances. For development and benchmarking, the quantum part of the algorithm is executed on the Qiskit quantum simulator. For experimentation and benchmarking, it is run on the publicly available \textsc{ibmq} Melbourne \cite{ibmq-specs} device, featuring 15 qubits.
This device is also used in recent research, including \cite{piveteau2021quasiprobability}.
Our implementation can be executed on both instances without modification of the code\footnote{The source code is available at: \url{https://github.com/macemoth/QuantumQueryOptimization}}.

The main difference between quantum simulators and real quantum devices is that the simulators produce results without noise. This means that on a simulator, one can work without error correction. 
With current hardware, up to 61 qubits can be simulated using supercomputers \cite{chic61qubits}, with desktop computers, quantum algorithms below 30 qubits are realistic \cite{Qiskit}.
As current quantum systems feature a comparable amount of qubits, the size of problems our algorithm can tackle is restricted by both quantum simulators and real quantum devices.


\subsection{Bootstrapping the Classical Optimization Part with a FOURIER Strategy}
When $\mathcal{F}_Q$ is constructed in Algorithm \ref{algo:lowlevel}, initial values for parameters $\gamma$ and $\beta$ need to be set to start the classical optimization. Usually those values are set randomly. Initializing $\gamma$ and $\beta$ at random has multiple disadvantages. First, the randomly generated values often lead the optimizer to converge against local optima and therefore the circuit returning sub optimal solutions. Second, the more repetitions of $\mathcal{F}_Q$ are applied to the circuit, the more parameters need to be initialized. Hence finding good initial parameters through random initialization gets exponentially more difficult.

\begin{figure}
    \centering
    \includegraphics[width=0.7\linewidth]{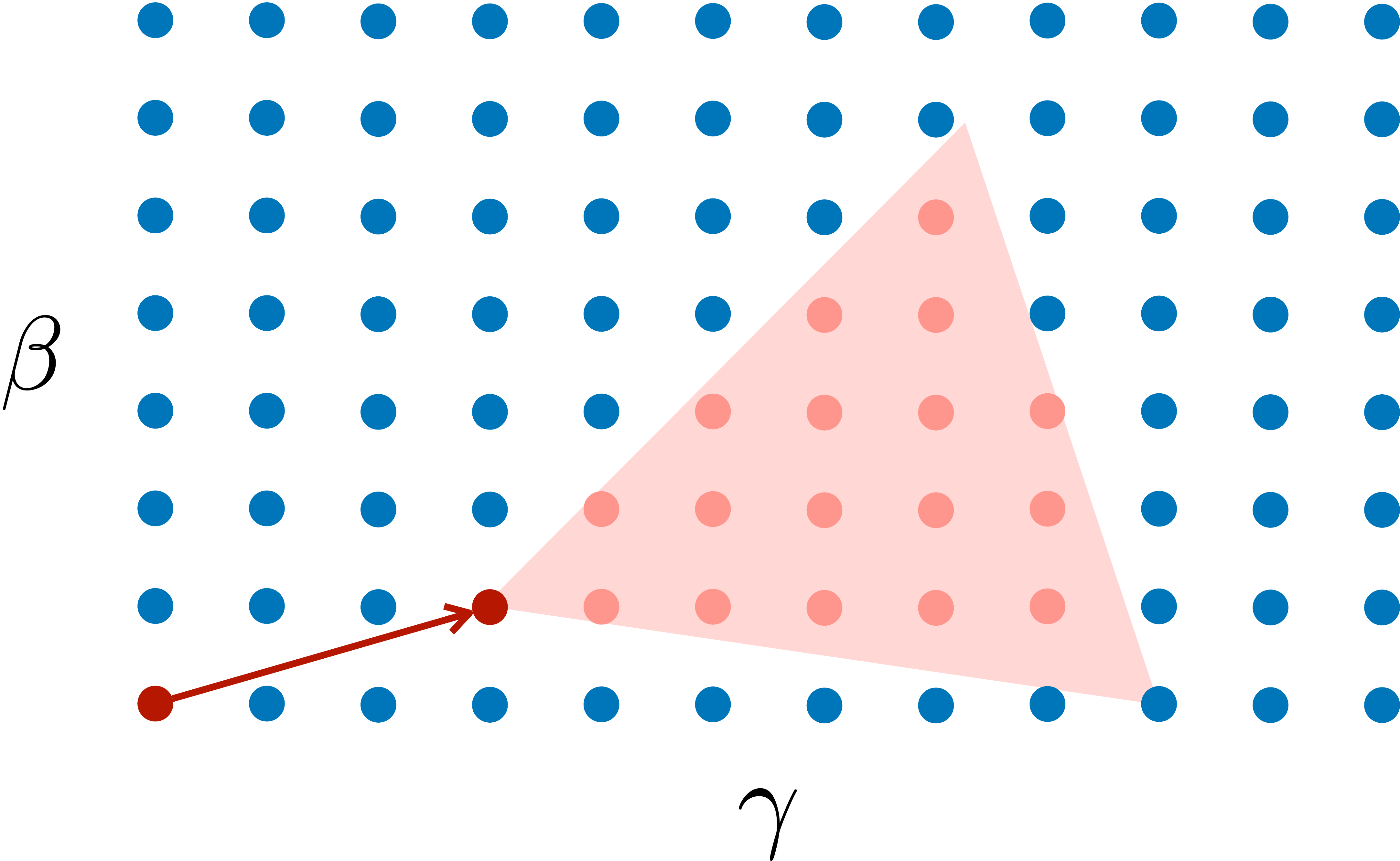}
    \caption[FOURIER strategy]{The FOURIER strategy restricts the parameter search space (light red) based on previous parameters (red).}
    \label{fig:fourier_lattice}
\end{figure}

We base our experiments on the FOURIER heuristic strategy introduced in \cite{zhou2020quantum} which seems to perform very well in finding good initial parameters for $\gamma$ and $\beta$. Instead of randomly initializing $\gamma$ and $\beta$ for every repetition of $\mathcal{F}$, the FOURIER strategy advises to calculate new values for all $\gamma_i$ and $\beta_i$ through the following transformation based on previously optimized values:

\begin{align*}
    \gamma_i &= \sum_{k=1}^p u_k \; \sin \left[ \left( k - \frac{1}{2} \right) \left( i - \frac{1}{2} \right) \frac{\pi}{p} \right], \\
    \beta_i &= \sum_{k=1}^p v_k \; \cos \left[ \left( k - \frac{1}{2} \right) \left( i - \frac{1}{2} \right) \frac{\pi}{p} \right]
\end{align*}

where the parameters $(u, v) \in \mathbb{R}^{2p}$ correspond to the parameters $(\gamma, \beta) \in \mathbb{R}^{2p}$. With this strategy only the initial parameters $\gamma_1$ and $\beta_1$ are determined randomly. Because of the rotational symmetry of qubit states the initial parameter guess can be restricted to $\beta_1, \gamma_1 \in [-\frac{\pi}{2}, \frac{\pi}{2})$. The schematic procedure for FOURIER is rendered in Figure \ref{fig:fourier_lattice}.

\begin{figure}
    \centering
    \includegraphics[width=0.7\linewidth]{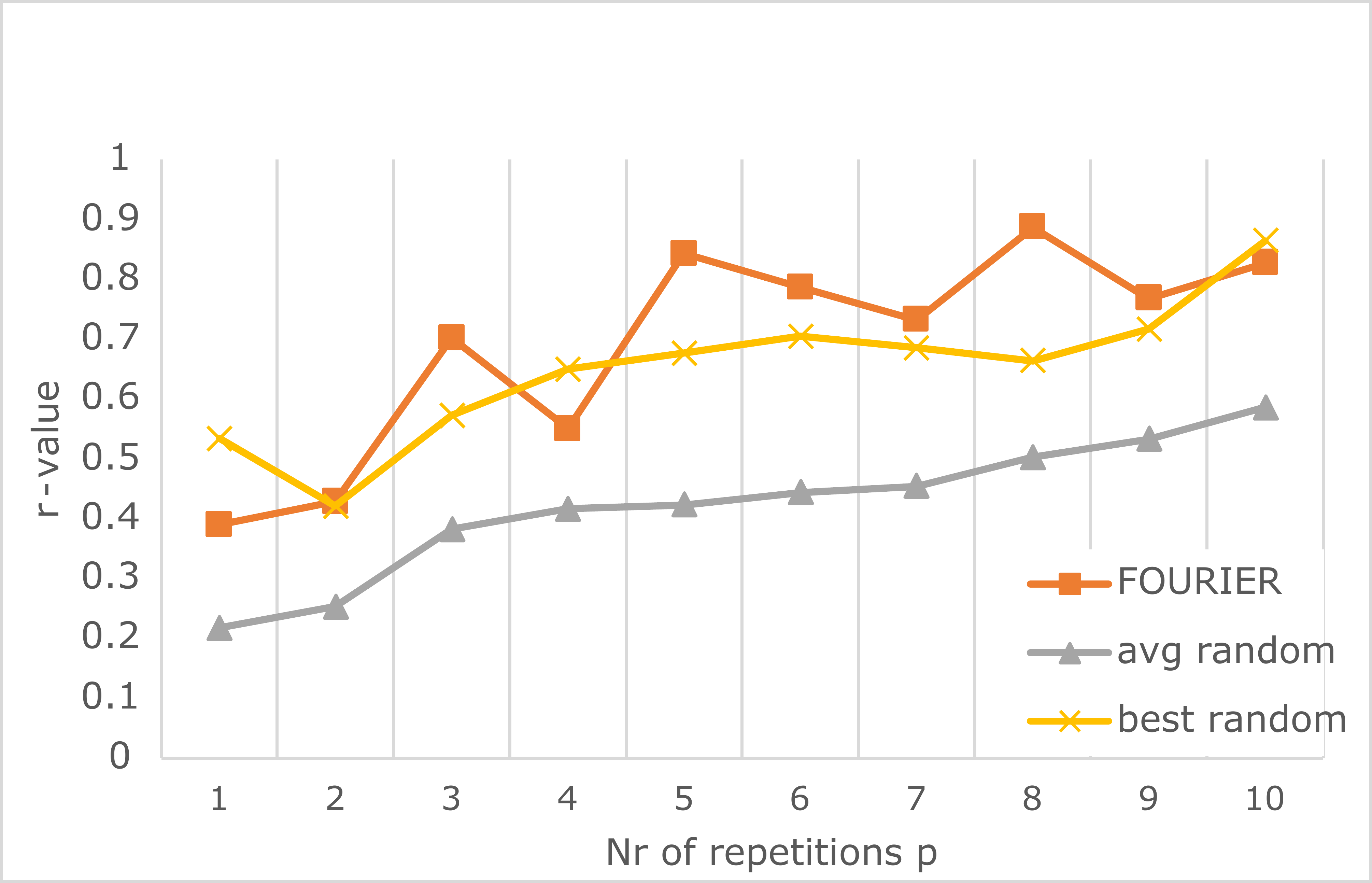}
    \caption[Comparison between the FOURIER strategy and randomly initialized parametersets]{Comparison between the FOURIER strategy and randomly initialized parameter sets for $\gamma$ and $\beta$. As can be seen, the FOURIER performs equally as the best of the randomly initialized parameters. The y-axis shows the approximation ratio $r$.}
    \label{fig:FOURIER_performance}
\end{figure}

To verify the results presented in \cite{zhou2020quantum} we compare the performance of the QAOA algorithm with parameters found through the FOURIER strategy against randomly initialized parameters. For the experiment we first generate a random multiple-query optimization problem of 4 qubits, i.e. 2 queries with 2 plans each. On that problem we apply the QAOA with different numbers of repetitions of $\mathcal{F}$. We perform 30 runs of the QAOA with FOURIER as well as with randomly initialized parameters for each number of repetitions of $\mathcal{F}$ on the simulator. The result can be seen in Figure \ref{fig:FOURIER_performance}. The average FOURIER strategy performs as well as the best randomly found parameters and is therefore our algorithm of choice for determining the parameters of the further experiments.  

\subsection{Applying Various Classical Optimization Approaches}
\begin{figure}
    \centering
    \includegraphics[width=0.7\linewidth]{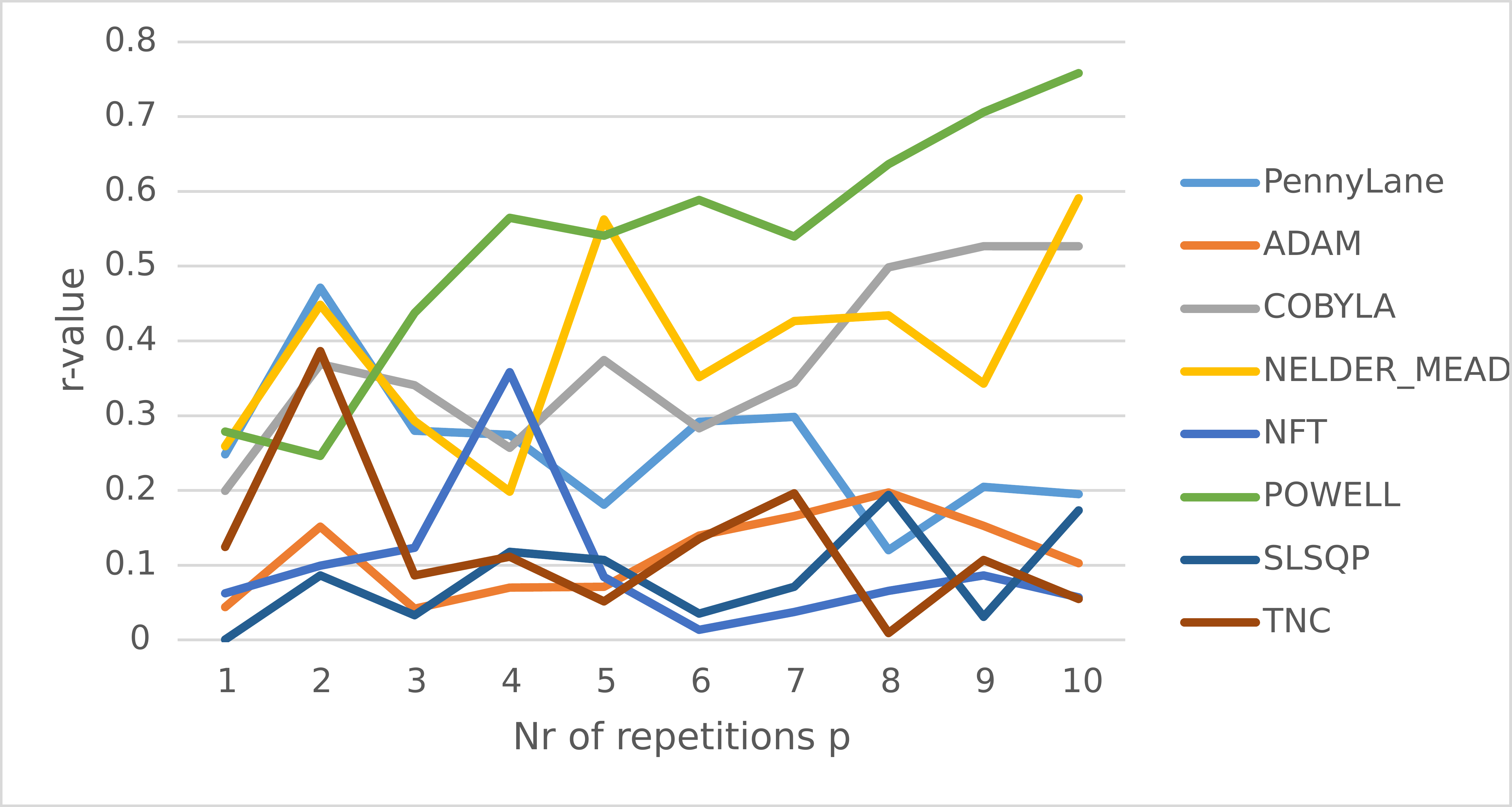}
    \caption[Comparison of the optimizers provided by the Qiskit and PennyLane]{Comparison of the optimizers provided by the Qiskit and PennyLane library on 30 different, randomly generated 4 qubit problem instances. The figure shows the average performance of the optimizer with regard to the approximation ratio $r$. As the number of repetitions of $\mathcal{F}$ increases, so does the performance of POWELL. Besides POWELL also NELDERMEAD and COBYLA seem to perform well compared to the other optimizers.}
    \label{fig:optimizer_comparison}
\end{figure}

Classical optimization arises to be the critical part of the QAOA because of the challenge to avoid local optima. To get around this problem, we evaluate the \textit{performance of the different optimizers} provided by the Qiskit library on the simulator \cite{Qiskit}. Additionally to the classical optimizer provided by Qiskit, we also use the PennyLane library with its quantum gradient descent optimizer \cite{Qiskit, bergholm2018pennylane}. Figure \ref{fig:optimizer_comparison} shows the performance of the different optimizers with respect to the number of applications of $\mathcal{F}$ in the circuit. All experiments are executed on the Qiskit simulator. As can be seen, for certain optimizers the quality of the solution increases with the number of repetitions as the theory suggests. Because of the steady increase and slight edge of performance of the POWELL optimizer \cite{Qiskit}, the further experiments are performed using this optimizer.

\subsection{Hybrid Classical-Quantum Optimization}

In this section we analyze the end-to-end query optimization problem using a hybrid classical-quantum optimization algorithm described by Algorithm 2. First we evaluate how to find the optimal parameters $\gamma$ and $\beta$, i.e. the classical part of Algorithm 2. Afterwards, we evaluate how to find the optimal number of repetitions of $F_Q$, i.e. the quantum part of Algorithm 2. Again, the classical part is run on the simulator while the quantum part is executed on the \textsc{ibmq} Melbourne quantum computer.

\subsubsection{Classical Part: Finding Optimal Parameters for $\gamma$ and $\beta$}
We first analyze the "classical part" of Algorithm 2, i.e. to find the optimal parameter setting of $\gamma$ and $\beta$. Note that finding the optimal parameters $\gamma$ and $\beta$ appears to be the main difficulty to apply the QAOA successfully on quantum computers \cite{farhi2014quantum, zhou2020quantum, harrigan2021quantum}. To get to the bottom of the problem, we try to determine what makes good $\gamma$ and $\beta$. Because this step happens with a classical optimizer e.g. gradient descent or other non-linear optimizers in a non-convex cost landscape, the optimizer can severely fail to find a good solution \cite{zhou2020quantum}.
Therefore, restricting the search space to good candidates for $\gamma$ and $\beta$ can drastically enhance the solution quality and the time consumed by classical optimizers. 

To evaluate whether there exist patterns in good candidates, we record the parameters which lead to right solutions for different problem sizes. We expect a steady increase in value for the parameter $\gamma_i$ compared to $\gamma_j$ with $i>j$ and a decrease in value for the parameter $\beta_i$ compared to $\beta_j$ with $i>j$. This behavior is predicted by the adiabatic principle and shown by \cite{zhou2020quantum}. 
\begin{figure}[h]
    \centering
    \begin{subfigure}[b]{0.4\textwidth}
        \centering
        \includegraphics[width=\textwidth]{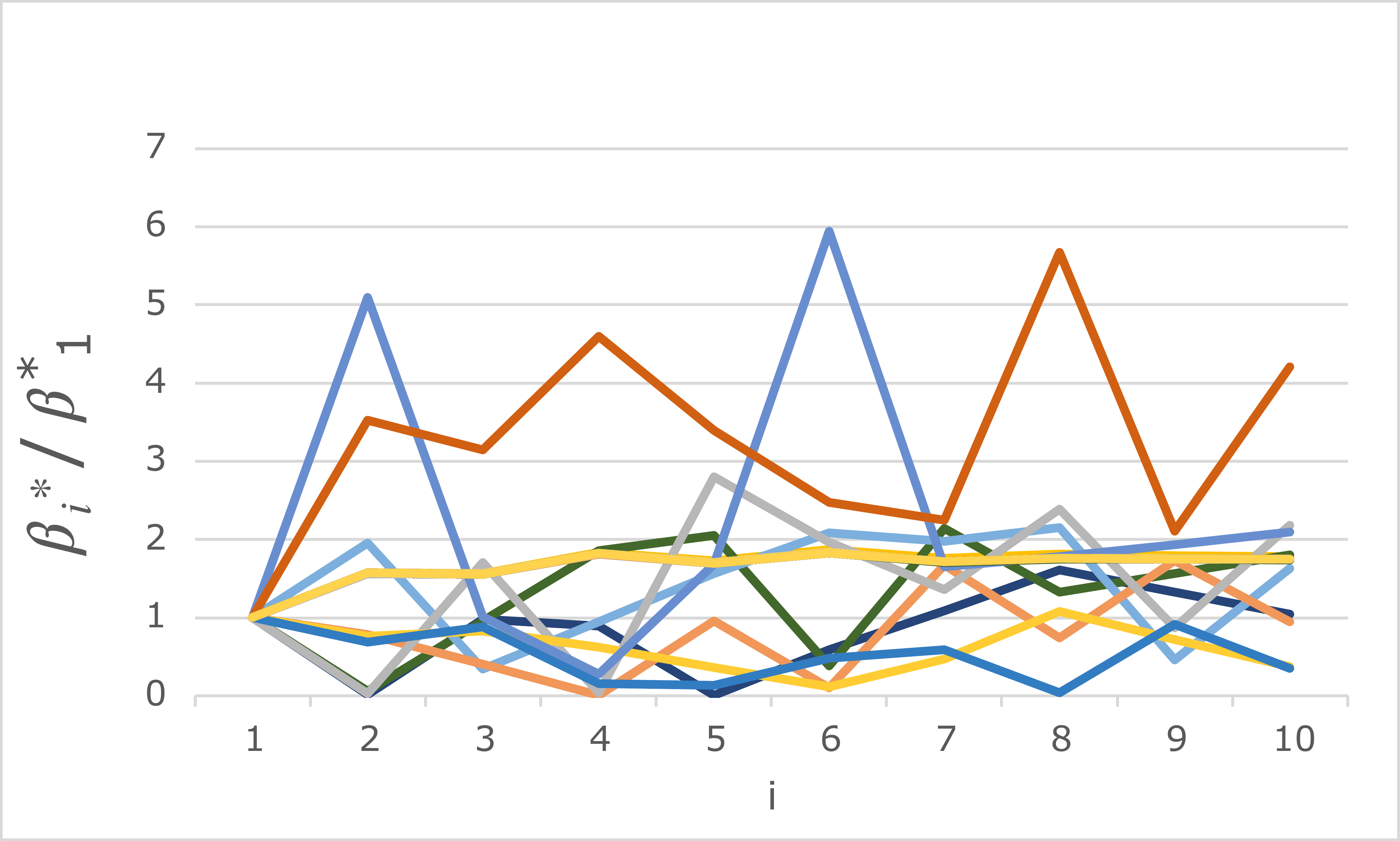}
        \caption{Correlation of optimal parameters $\beta_i$ for various problems of size 4 Qubits}
    \end{subfigure}
    \begin{subfigure}[b]{0.4\textwidth}
        \centering
        \includegraphics[width=\textwidth]{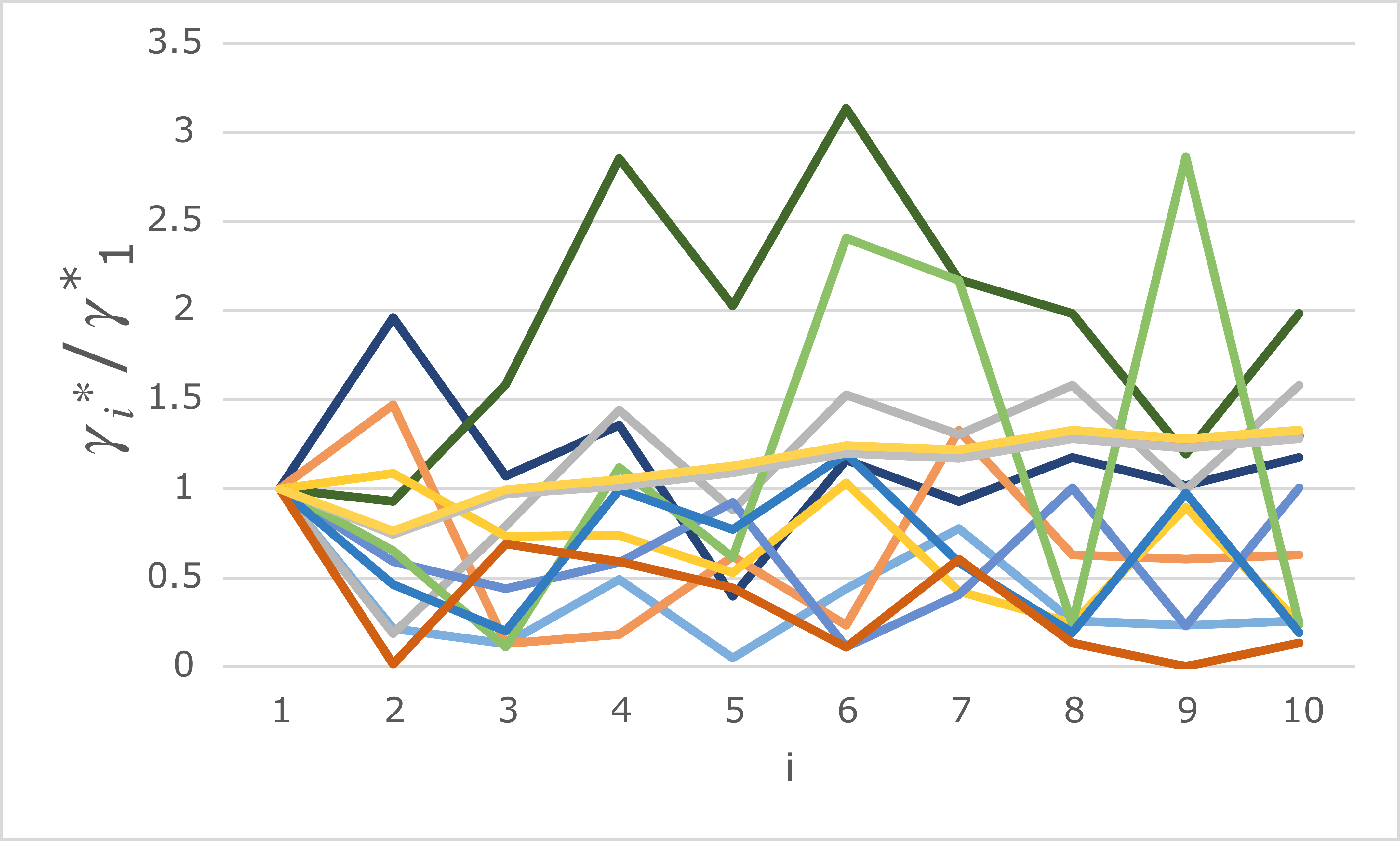}
        \caption{correlation of optimal parameters $\gamma_i$ for various problems of size 4 Qubits}
    \end{subfigure}
    \caption[The correlation between parameters of different $\mathcal{F}$ of a circuit]{The correlation between parameters of different $\mathcal{F}$s of a circuit, visualized by plotting the optimal parameters of 35 correctly solved instances of different, randomly generated MQO problems with the QAOA and various optimizers. One line represents all parameters of the same circuit. For instance $\beta_i$ is the parameter $\beta$ of $\mathcal{F}_i$. To make the generated data comparable to the eye, the initial values are normalized and the subsequent values are also divided by the initial value.}
    \label{fig:beta_gamma}
\end{figure}

 Although the application of the FOURIER strategy always resets the initial parameters in a similar way, the classical optimization part finds good solutions scattered all over the search space. As can be seen in Figure \ref{fig:beta_gamma}, there is no obvious pattern in the parameter setting of $\gamma^*$ and $\beta^*$, such that the QAOA can extract the optimal solution. Hence, the optimal parameter setting  must be determined empirically for each problem.

\subsubsection{Quantum Part: Finding Optimal Repetitions of\, $\mathcal{F_Q}$}
Next, we  analyze the "quantum part" of Algorithm 2, i.e. to find the optimal number of repetitions of $F_Q$ depending on the size of our multiple-query optimization problem. In particular, we vary the number of queries as well as the number of query plans and study the probabilities of finding the optimal solution, i.e. the optimal query plan, for various problem sizes.

The QAOA algorithm is currently not well understood beyond $p = 1$ repetitions of $\mathcal{F_Q}$. To investigate the behavior and necessity of more than 1 repetition for larger problem sizes, we apply the QAOA on various randomly generated MQO problems up to problem sizes of 14 plans (resulting in the same number of qubits). In theory, and empirically proven on the simulator in our experiments before, more applications of $\mathcal{F_Q}$ should lead to a better approximation of the minimum cost but at the same time deepening the circuit and increasing the gate induced error \cite{zhou2020quantum,harrigan2021quantum}. 

We perform this experiment on the \textsc{ibmq} Melbourne quantum computer where we evaluate the accuracy of the QAOA on 50 different problems on each problem size.
\begin{figure}
    \includegraphics[width=0.5\textwidth]{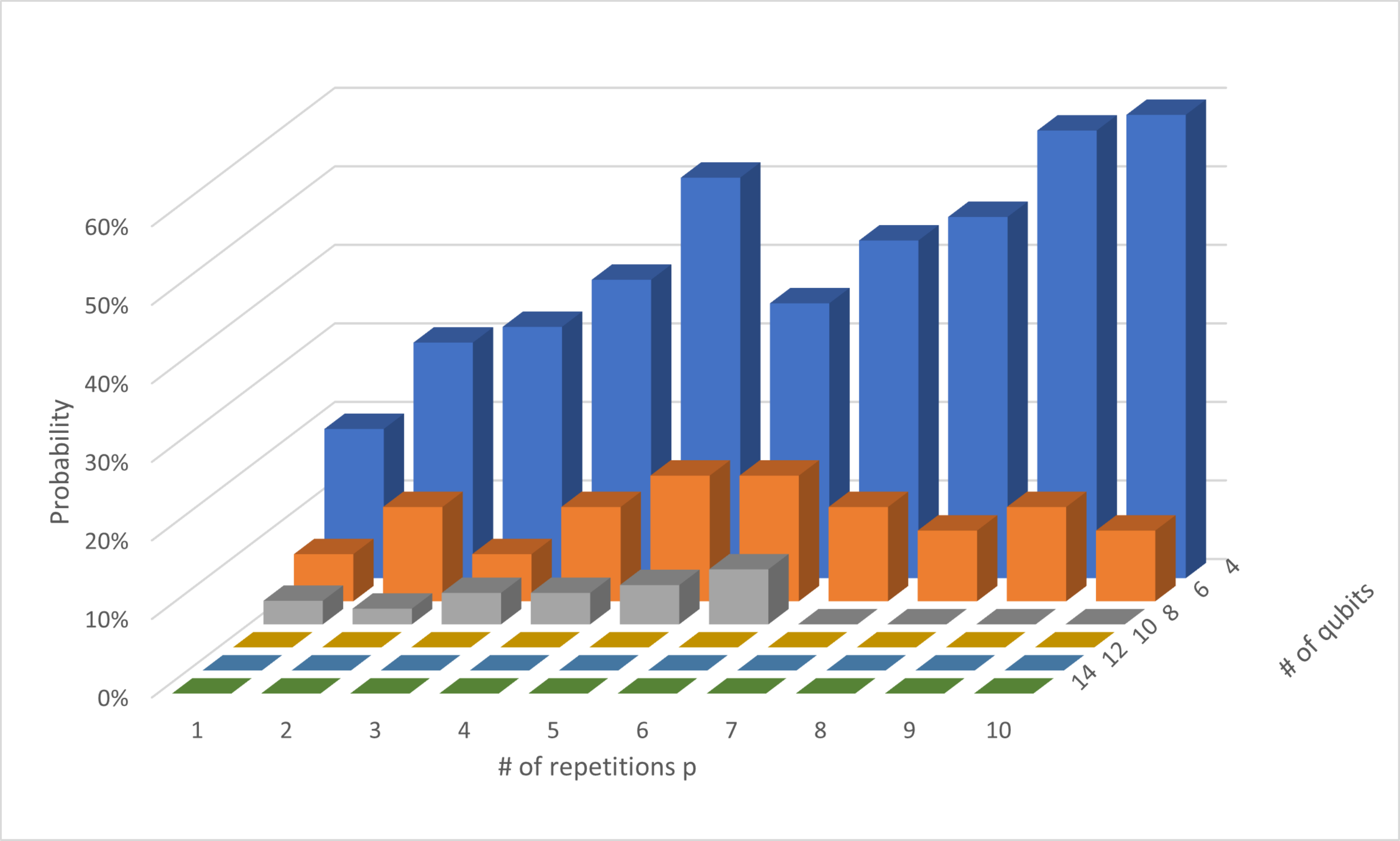}
    \caption[Accuracy of measuring the correct solution for different problem sizes and values of p]{The accuracy of measuring the correct solution for problems of size 4 to 14 qubits with 1 up to 10 repetitions of $\mathcal{F_Q}$ in the circuit. The accuracy peaks at 59\% on problem instances with 4 qubits i.e. 2 queries with 2 plans each. On problems of size 6 qubits the accuracy ceils by 16\% with 6 repetitions of $\mathcal{F}$. Problem instances of size larger than 8 qubits have a 0\% accuracy on the \textsc{ibmq} Melbourne in our experiment.}
    \label{fig:qubitsvsp}
\end{figure}

Figure \ref{fig:qubitsvsp} shows the probability of finding the optimal solution as a function of the number of qubits, i.e. problem size, and the number of repetitions $p$ of $F_Q$. We measured the results on problems of size 4 to 14 qubits and with 1 to 10 repetitions of $\mathcal{F_Q}$ in the circuit. Our experimental results show that for a problem size of 4 qubits, i.e. 2 queries with 2 query plans each, the probability of finding the optimal solution is 59\%. For a problem size of 6 qubits, with 4 repetitions of $\mathcal{F_Q}$, the optimal solution is found with a probability of 16\%. For problem sizes of 8 qubits and above, the probability of finding the optimal solution converges to zero. In other words, deeper circuits can currently not be processed by the \textsc{ibmq} Melbourne device.


\subsection{Experimental Algorithm Run Times}
Each run is concluded by measuring the state of the qubits, which is a stochastic procedure. For statistics, the quantum circuit is executed and then measured multiple times, where each cycle is called a \textit{shot}. Typically, thousand to ten thousand shots are conducted \cite{guerreschi2019qaoa}.
Table \ref{tab:runtimes} shows the minimum, median and maximum run times for four differently sized problems. For all experiments, five repetitions of $\mathcal{F}$, that is, $p=5$ were used on the \textsc{ibmq} Melbourne device.

\begin{table}[h]
\centering
\caption{Run times of quantum circuits (shots) for different problems with the number of queries $Q$ and number of plans per query $P$ in milliseconds.}
    \begin{tabular}{l l l l l l}
        & & & \multicolumn{3}{c}{Run time in ms} \\ \hline
        & \textbf{$Q$} & \textbf{$P$} & \textbf{Min.} & \textbf{Med.} & \textbf{Max.}  \\ \hline
        \multirow{2}{*}{14 Qbits} & 7 & 2 & 14.2 & 28.0 & 35.3 \\
        & 2 & 7 & 23.8 & 25.0 & 30.9 \\ \hline
        \multirow{2}{*}{7 Qbits} & 4 & 2 & 14.3 & 15.1 & 15.2 \\
        & 2 & 4 & 14.6 & 15.2 & 15.7
    \end{tabular}
    \label{tab:runtimes}
\end{table}

The run times mainly depend on the number of qubits used, which are determined by the number of plans $\times$ the number of queries. If there are many alternative plans for each query (i.e. $P$ is large), the execution time increases slightly.
Although the doubling of qubits used leads to doubling in run times, this correlation might be coincidental. The time complexity is analyzed in more detail at the end of this section.
Generally, the execution time is not only determined by the amount of gates, but as follows: (time to prepare the qubit's state) + (time per gate) $\times$ (amount of gates) + (time to measure) \cite{guerreschi2019qaoa}.

\subsection{Comparison with Competing Approaches}
For the MQO problem \cite{sellis1988multiple}, a variety of approaches have been suggested. Early methods were based on graphs, using algorithms based on A* and branch-and-bound techniques \cite{grant1982optimizing, sellis1988multiple}.
Since then, integer linear programming \cite{dokeroglu2014integer} and genetic algorithms \cite{bayir2006genetic} have been proposed, improving prior methods.
More recent approaches employ dynamic programming, while considering multiple cost metrics \cite{trummer2016multi}.
In case of the more general class of query optimization problems, learning algorithms have successfully been applied \cite{stillger2001leo}. Modern approaches extend this approach with the usage of deep learning methods \cite{marcus2019neo,heitz2019join}.

Whilst almost all of the existing approaches rely on classical computing architecture, to the best of our knowledge, there is only one quantum-based solution to address the MQO suggested by \textit{Trummer and Koch} \cite{trummer2015dwave}. In the following paragraph, we compare these two approaches in more detail.

In their paper, Trummer and Koch utilize a D-Wave quantum annealer with more than 1000 qubits \cite{trummer2015dwave}, while we employ a gate-based quantum computer with 15 qubits \cite{ibmq-specs}. The qubit disparity between quantum annealers and gate-based quantum computers persists to this day, with quantum annealers featuring almost two orders of magnitude more qubits than gate-based quantum computers \cite{dwave, ibmq-manhattan}.
Clearly, the D-Wave device's superiority in qubits allows it to solve larger problems with as many as 1074 queries with two plans each, or 540 queries with 5 plans each, as shown in Table \ref{tab:dwave_comparison}.

\begin{table*}
    \centering
    \caption[Comparison of the approaches by Trummer and Koch \cite{trummer2016multi} and ours]{Comparison of the approaches by Trummer and Koch \cite{trummer2016multi} and ours. Although the D-Wave quantum device can tackle significantly larger problems, there are situations it can only use a fraction of its qubits.}
    \label{tab:dwave_comparison}
    \begin{tabular}{l l l}
        & \textbf{Trummer and Koch} & \textbf{This work}  \\ \hline
        Max. \# of queries & 537 queries, 2 plans & 7 queries, 2 plans \\
        Qubits used & 1074 / 1097 (98\%) & 14 / 14 (100\%) \\ \hline
        Max. \# of plans & 108 queries, 5 plans & 2 queries, 7 plans \\
        Qubits used & 540 / 1097 (49\%) & 14 / 14 (100\%) \\
    \end{tabular}
\end{table*}

\begin{figure}
    \centering
\begin{tikzpicture}[shorten >=0.1pt,auto,node distance=3cm,
        thick,node/.style={circle,draw,minimum size=0.4cm,inner sep=0pt]}]

		\node [node, style={fill=blue!20}] (0) at (0, 1.5) {1};
		\node [node, style={fill=blue!80}] (1) at (0.75, 0) {4};
		\node [node, style={fill=blue!40}] (2) at (0, 1) {2};
		\node [node, style={fill=blue!60}] (3) at (0, 0.5) {3};
		\node [node, style={fill=blue!80}] (4) at (0, 0) {4};
		\node [node, style={fill=blue!60}] (5) at (0.75, 0.5) {3};
		\node [node, style={fill=blue!40}] (6) at (0.75, 1) {2};
		\node [node, style={fill=blue!20}] (7) at (0.75, 1.5) {1};
		\node [node, style={fill=blue!20}] (16) at (0, -1) {1};
		\node [node, style={fill=green!80}] (17) at (0.75, -2.5) {8};
		\node [node, style={fill=blue!40}] (18) at (0, -1.5) {2};
		\node [node, style={fill=blue!60}] (19) at (0, -2) {3};
		\node [node, style={fill=blue!80}] (20) at (0, -2.5) {4};
		\node [node, style={fill=green!60}] (21) at (0.75, -2) {7};
		\node [node, style={fill=green!40}] (22) at (0.75, -1.5) {6};
		\node [node, style={fill=green!20}] (23) at (0.75, -1) {5};
		\node [node, style={fill=green!20}] (24) at (1.75, -1) {5};
		\node [node, style={fill=green!80}] (25) at (2.5, -2.5) {8};
		\node [node, style={fill=green!40}] (26) at (1.75, -1.5) {6};
		\node [node, style={fill=green!60}] (27) at (1.75, -2) {7};
		\node [node, style={fill=green!80}] (28) at (1.75, -2.5) {8};
		\node [node, style={fill=green!60}] (29) at (2.5, -2) {7};
		\node [node, style={fill=green!40}] (30) at (2.5, -1.5) {6};
		\node [node, style={fill=green!20}] (31) at (2.5, -1) {5};

		\draw (0) to (7);
		\draw (0) to (6);
		\draw (0) to (5);
		\draw (0) to (1);
		\draw (2) to (7);
		\draw (2) to (6);
		\draw (2) to (5);
		\draw (2) to (1);
		\draw (3) to (7);
		\draw (3) to (6);
		\draw (3) to (5);
		\draw (3) to (1);
		\draw (4) to (7);
		\draw (4) to (6);
		\draw (4) to (5);
		\draw (4) to (1);
		\draw (16) to (23);
		\draw (16) to (22);
		\draw (16) to (21);
		\draw (16) to (17);
		\draw (18) to (23);
		\draw (18) to (22);
		\draw (18) to (21);
		\draw (18) to (17);
		\draw (19) to (23);
		\draw (19) to (22);
		\draw (19) to (21);
		\draw (19) to (17);
		\draw (20) to (23);
		\draw (20) to (22);
		\draw (20) to (21);
		\draw (20) to (17);
		\draw (24) to (31);
		\draw (24) to (30);
		\draw (24) to (29);
		\draw (24) to (25);
		\draw (26) to (31);
		\draw (26) to (30);
		\draw (26) to (29);
		\draw (26) to (25);
		\draw (27) to (31);
		\draw (27) to (30);
		\draw (27) to (29);
		\draw (27) to (25);
		\draw (28) to (31);
		\draw (28) to (30);
		\draw (28) to (29);
		\draw (28) to (25);
		\draw [bend left=40, looseness=0.50] (23) to (31);
		\draw [bend left=40, looseness=0.50] (22) to (30);
		\draw [bend left=40, looseness=0.50] (21) to (29);
		\draw [bend left=40, looseness=0.50] (17) to (25);
		\draw [bend right=25] (0) to (16);
		\draw [bend right=25] (2) to (18);
		\draw [bend right=25] (3) to (19);
		\draw [bend right=25] (4) to (20);
\end{tikzpicture}
    \caption[Chimera graph of the D-Wave quantum annealer]{Chimera graph representing the qubits of the D-Wave quantum annealer \cite{trummer2015dwave}. The vertices denote qubits with the variable number inside and the edges denote connections among qubits. As the graph is not fully connected, variables must be mapped to multiple qubits, ensuring that all variables are fully connected among each other.}
    \label{fig:chimera}
\end{figure}
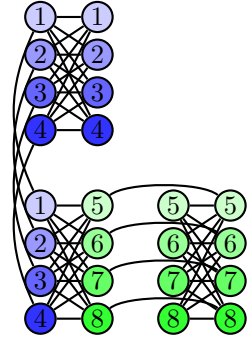

Similarly to our approach, Trummer and Koch assign plans to qubits that are either selected or not to form the solution to the MQO. For this, they map plans as variables to the physical qubits, called physical mapping \cite{trummer2015dwave}. The physical layout of the qubits on the D-Wave quantum annealer is represented as Chimera graph \cite{mcgeoch2013experimental} as in Figure \ref{fig:chimera}.

To model cost savings, which become active when certain pairs of qubits have been selected for the solution, qubits have to interact with each other (this is, they are entangled). For this interaction, plan-encoding qubits have to be connected to each other. As the Chimera graph is not fully connected, it is necessary to \textit{encode a plan in multiple qubits} to ensure that the necessary interactions can be facilitated. An example for this encoding is the TRIAD pattern (see Fig. \ref{fig:chimera}) \cite{trummer2015dwave}, where a total of eight variables are mapped to 24 qubits to become fully connected.

This mapping introduces two restrictions. First, the physical \textit{mapping introduces an overhead}, in particular with problems that require strong interaction. An example of such a problem is the MQO with many plans per query. With the requirement to only choose one plan per query, all plans within that query must be connected, augmenting the degree of interaction. As a consequence, with a higher number of plans per query, around half of the quantum annealer's qubits are required for the physical mapping, as shown in Table \ref{tab:dwave_comparison}.
The second restriction is that the physical mapping with the minimal number of qubits is itself an \textit{NP-hard problem} \cite{klymko2014adiabatic}.

Although \textit{our approach} is heavily limited by the number of qubits of gate-based quantum devices, it \textit{does not suffer both of these restrictions}, as all of the available qubits can be used to encode plans and the physical mapping is a one-to-one mapping from a variable to a qubit.

Both our approach on QAOA as well as quantum annealing are founded on the quantum-mechanical adiabatic principle \cite{farhi2014quantum,farhi2000quantum,zhou2020quantum}.
In quantum annealing, the minimum spectral gap $\Delta_{\min}$, informally described as a property of the problem-encoding function $\mathcal{F}$, determines how much time $T$ is required for the adiabatic evolution \cite{zhou2020quantum} (we use the conventions and units of this work). The relationship is given by

\begin{equation}
    T = \frac{1}{\Delta_{\min}^2}.
\end{equation}

\begin{figure}[h]
    \centering
    \includegraphics[width=0.6\linewidth]{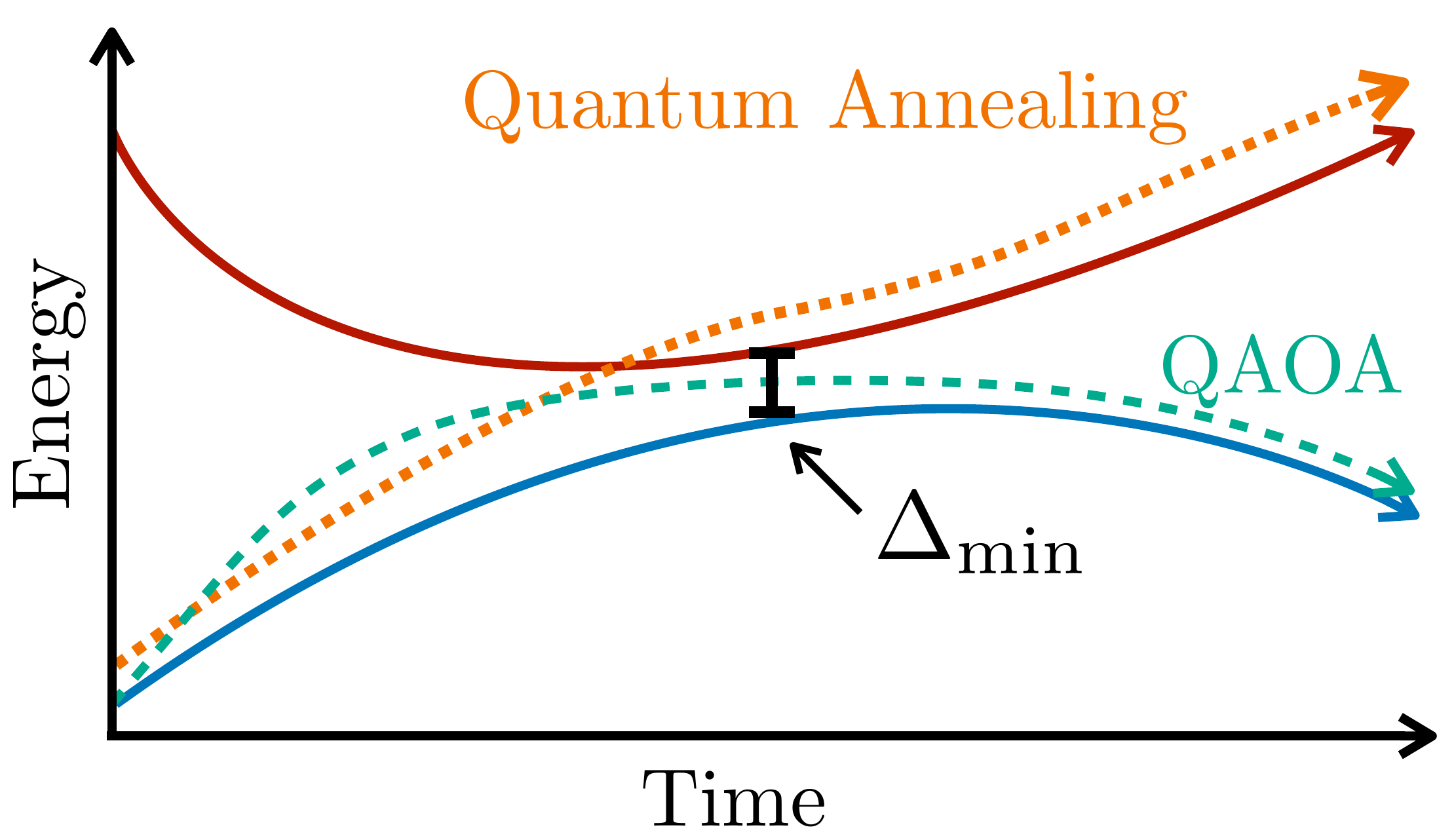}
    \caption[The minimum spectral gap of quantum annealing]{The minimum spectral gap $\Delta_{\min}$ visualized between the blue and red lines as energy states. While quantum annealing is susceptible to leave its path with small $\Delta_{\min}$ and to follow the red energy state, the QAOA can overcome this gap and remains close to the desired blue energy state. $s = t/T$ denotes the state of the evolution with the total annealing time $T$.}
    \label{fig:minimumgap}
\end{figure}

Zhou et al. \cite{zhou2020quantum} report that some graph-based problems ("hard problems") exhibit very small minimum spectral gaps, thus requiring a very long time to evolve, as described by the relationship above. When evolved too quickly, the quantum annealing algorithm can leave it's "path", and thus produces a completely different result \cite{farhi2000quantum}, as illustrated in Figure \ref{fig:minimumgap}.

The time required to evolve would in practice render quantum annealing ineffective for those problems.
Zhou et al. further report that QAOA can overcome this limitation, as it adjusts its parameters during the optimization procedure. Hence, QAOA can conclude the adiabatic evolution much faster and solve problems that can not effectively be solved by quantum annealing.
This resistance makes QAOA more robust in tackling these hard problems.

\begin{figure}
    \centering
    \includegraphics[width=0.7\linewidth]{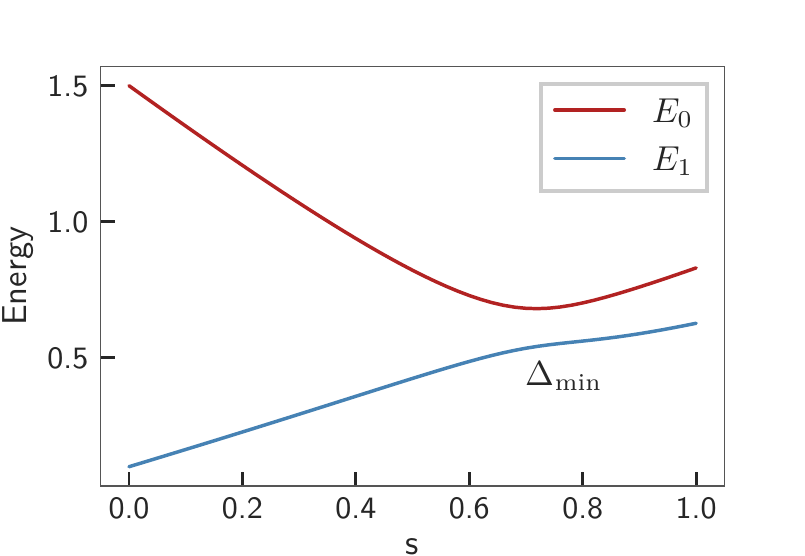}
    \caption[The minimum spectral gap of the example instance]{The minimum spectral gap $\Delta_{\min} = 0.134$ between the energy levels $E_0$ and $E_1$ of the example MQO.}
    \label{fig:minimumgap_example}
\end{figure}

Zhou et al. classify gaps $\Delta_{\min} < 10^{-3}$ as very small \cite{zhou2020quantum} and thus problematic for quantum annealing. Existing approaches typically compute the minimal gap for random problem instances \cite{zhou2020quantum} or estimate it for particular problems \cite{farhi2000quantum}. Farhi et al. state that they are in general unable to estimate the gap. Choi \cite{choi2020effects} examines the effects of the problem\footnote{More specifically, the problem Hamiltonian.} on the gap analytically. Choi notes that the relation between the problem encoding (introduced as $\mathcal{F}_Q$) and the driver gates (see Section \ref{ssec:implementing}) plays a significant role for the minimum gap.
For general search problems, Albash et al. \cite{albash2018adiabatic} indicate an inverse scaling of the minimum gap with the number of qubits (i.e. problem size), identifying the gap (-scaling) as a problem for quantum annealing.
For the MQO, we numerically evaluate the minimum gap for several problem instances (see Table \ref{tab:min_gaps}).
The energy states of the example problem introduced in Section \ref{sec:introduction} are shown in Figure \ref{fig:minimumgap_example}.

We observe that with a larger number of plans per query $P$, $\Delta_{\min}$ typically increases. Even when increasing the number of qubits (this is, $P \cdot Q$), the minimum gap grows. This effect is different than reported for many problems in the literature.
We suggest that the benevolent gap scaling is likely due to the MQO's problem structure, where the amount of plans per query $P$ contributes strongly to the amount of qubits but has a positive influence on the minimum gap.

By constructing a problem with small $P$ but large $Q$, we indeed find that the minimum gap becomes smaller.
Similarly to Zhou et al., we find a MQO problem instance that has a extremely small gap $\Delta_{\min} \approx 3.33 \cdot 10^{-16}$ with plan cost of 8, 42, 49,  3 and savings $s_{1,3} = -15$, $s_{1,4} = -12$, $s_{2,3} = -5$ and $s_{2,4} = -14$. We suspect that the very small gap partially stems from the large savings, and are able to reproduce this problem to some degree by introducing large savings. However, we are not able to find a pattern that routinely leads to very small gaps.

This coincides with the observations of most authors, that very small gaps are observed for special instances found at random rather than by construction.
Based on the random problem instances analyzed, we suppose that there may be real MQO instances that exhibit very small minimum gaps that are indeed problematic for quantum annealers and therefore better tractable by our approach instead of quantum annealing.

\begin{table}[]
    \centering
    \begin{tabular}{c c | c c}
    \multicolumn{2}{c}{\textbf{Plans}} & \multicolumn{2}{c}{\textbf{Qubits}} \\
        $P$ & $\Delta_{\min}$ & \# of qubits & $\Delta_{\min}$ \\ \hline
        2 & 0.134 & 4 & 0.134 \\
        3 & 0.180 & 6 & 0.076\\
        4 & 0.169 & 8 & 0.140\\
        5 & 0.262 & 10 & 0.213\\
    \end{tabular}
    \caption{Minimum gaps for different numbers of plans $P$ with fixed $Q=2$ (left) and for different number of qubits (right). The instances on the RHS consist of different combinations of $Q$ and $P$ to satisfy $P \cdot Q$ = \# of qubits.}
    \label{tab:min_gaps}
\end{table}

\subsection{Scaling the Problem}
Current quantum devices are severely limited by the number of qubits they feature. Moreover,  gate-based quantum computers, as used for this work, are additionally restricted because each operation introduces non-negligible error \cite{proctor2020demonstrating,zhou2020quantum}.
Quantum algorithms, such as Shor's algorithm \cite{shor1999polynomial} assume availability of fault-tolerant quantum computers comprising millions of qubits \cite{moll2018quantum}. Those computers cannot be constructed currently. In addition, algorithms designed to run on near-term quantum computers, such as QAOA \cite{farhi2014quantum}, can currently only be used to solve very small proof-of-concept problems.
Hence, solving such problems on current gate-based quantum computers gives only limited insight in the real-world use of quantum algorithms. Simulators remain to experimentally examine quantum algorithms merely beyond this insight\footnote{A notable exception is Arute et al.'s experiment, claiming to outperform a classical simulation \cite{google_quantum}.}.
Nevertheless, it remains to analyze quantum algorithms analytically. In the next paragraph we analyze our algorithm in terms of scaling, as the intention is to devise quantum algorithms that potentially solve problems faster than classical computers.

We provide a sketch of the computational complexity of our algorithm and compare it with the classical brute-force approach introduced in Section \ref{sec:background}.
Recall that for the MQO, we denote the number of queries by $Q$, the number of plans per query by $P$ and the total number of plans by $PQ$. For each of the $Q$ queries, one of the $P$ plans must be selected. With this requirement, only admissible solutions remain. 

\begin{centering}
\begin{tabular}{c c c c}
        $q_1$ & $q_2$ & $\ldots$ & $q_Q$ \\
        $\underbrace{\color{gray}{p_1}, \color{gray}{p_2}, \color{black}{p_3}}$ & $\underbrace{\color{gray}{p_4}, \color{black}{p_5}, \color{gray}{p_6}}$ & $\ldots$ & $\color{black}{p_{PQ-2}}, \color{gray}{p_{PQ-1}}, \color{gray}{p_{PQ}}$ \\
        $P$ poss. & $P$ poss. & \ldots & $P^Q$ possibilities
\end{tabular}\\
\medskip
\end{centering}

By brute force, $P^Q$ possibilities must be evaluated\footnote{If we included non-admissible solutions with no or more than one plan per query, $2^{PQ}$ evaluations would be required.}. Considering a constant time to evaluate the cost of each solution, the \textit{complexity of the brute force approach} is as follows:

\begin{equation}
    O(Brute Force) = O(P^Q)
\end{equation}

For the quantum approach, we focus on the quantum part and assume the complexity of the classic optimization as constant $O(1)$. Furthermore, we assume the complexity for formulating the MQO into a classical cost function, and into a quantum circuit to be constant.

The \textit{space complexity} of the quantum circuit can trivially be answered as $O(P)$ because our approach requires exactly one qubit for every plan of the MQO.
The \textit{time complexity} for quantum computations is typically determined by the depth of the quantum circuit \cite{guerreschi2019qaoa}. This is because the depth indicates the longest succession of gates that must be executed sequentially. The time required for a computation to complete is obtained by multiplying the amount of operations with the average time a quantum gate requires \cite{guerreschi2019qaoa}.

For the complexity sketch, we split the quantum cost function $\mathcal{F}_Q$ into its three sums $\mathcal{SQ}_1$ to $\mathcal{SQ}_3$.
$\mathcal{SQ}_1$ contains linear terms and only adds one gate to each qubit. Because these gates are executed simultaneously, the complexity is constant with $O(1)$.

$\mathcal{SQ}_2$ is a quadratic sum and appends a gate for every pair of plans that, when active, save costs. Although very untypical, every plan could potentially exhibit cost savings when selected with any other plan (e.g. when there is an intermediate result that can be reused for all queries). The number of all plan pairs is then $PQ(PQ-1) / 2$, resulting in a complexity of $O((PQ)^2)$. Because only one plan can be selected per query, plans for the same query could never form pairs. Because of this, we consider this estimate an upper bound.

Finally, $\mathcal{SQ}_3$ is another quadratic sum. It penalizes the selection of more than one plan per query. For this, it connects all $P$ plans inside a query with each other pairwisely by adding a gate to every pair. Similar to above, all plan pairs inside the query are $P(P-1) / 2$, contributing to a complexity of $O(P^2)$. These gates can be applied simultaneously for all queries, therefore, the number of queries is not decisive.

As the three components of $\mathcal{F}_Q$ are sequentially connected, the individual complexity sketches can be added. Because $\mathcal{F}_Q$ is run $I$ times in the optimization loop with the classical optimizer (which is assumed $O(1)$ here), the complexity of QAOA is 

\begin{equation}
O(QAOA) = O(I) \cdot (O(P) + O((PQ)^2) + O(P^2))
\end{equation}

After reformulation, we end up with the following complexity of QAOA:

\begin{equation}
O(QAOA)  = O(I\cdot (PQ)^2)
\end{equation}

Compared to the brute force complexity of $O(P^Q)$, this is a significant speedup. For comparison, Grover's algorithm for searching an unstructured list achieves a speedup from $O(n)$ to $O(\sqrt{n})$.
This suggests that the quantum part of the algorithm does not only work for very small problems, but also for real-world sized problems, as the number of gates grows moderately.

\begin{example}
In both classical and quantum computing, the amount of operations determines the time and feasibility for a computation \cite{nielsen2002quantum,guerreschi2019qaoa}.
In this example, we estimate and compare the amount of operations of the QAOA algorithm with the brute force approach for different problem sizes (see Table \ref{tab:optimization_complexity}).
For our quantum query optimization algorithm, we neglect the number of iterations of the optimization loop, as the same circuit is run in each iteration. Thus, the number of iterations does not determine the amount of quantum gates required.

\begin{table}
\caption{Amount of operations for solving a multiple query optimization problem as a function of the problem size, i.e. the number of queries $Q$ and the number of plans $P$. The classical approach is a brute force algorithm. Our approach is the hybrid quantum-classical approach QAOA.}
    \begin{tabular}{l l l}
        & \multicolumn{2}{c}{\# of operations} \\ \hline
        \textbf{Problem size}& \textbf{Brute Force} & \textbf{QAOA}  \\ \hline
        $Q = 2$, $P = 2$ & $4$ & $16$ \\
        $Q = 10$, $P = 10$ & $10^{10}$ & $10^{5}$ \\
        $Q = 100$, $P = 10$ & $10^{100}$ & $10^{6}$ \\
    \end{tabular}
    \label{tab:optimization_complexity}
\end{table}

With larger problem instances, our algorithm requires clearly fewer operations than a brute force algorithm.
Still, the two larger problem sizes could not be solved on a current gate-based quantum computer \cite{ibmq-specs}, and as described previously.
\end{example}

The complexity sketch above, however, neglects the non-trivial classical optimization procedure, which is analyzed in Zhou et al. \cite{zhou2020quantum} and from a more theoretical aspect in Farhi and Harrow \cite{farhi2019quantum}. In the same paper, Farhi and Harrow analyze the prospects of QAOA for near-term quantum devices and as candidate for a problem-solving technique superior to any classical approach.
Guerreschi and Mazuura \cite{guerreschi2019qaoa} argue that the necessary speedup with QAOA can be attained with hundreds of qubits. They also propose a complete estimate on the computational time of realistic QAOA applications.

\section{Conclusion} \label{sec:conclusion}
In this paper we have proposed a hybrid classical-quantum algorithm based on the Quantum Approximate Optimization Algorithm to find quasi-optimal solutions for the multiple query optimization problem, a classical problem in database optimization.
Our algorithm is designed for near-term gate-based quantum computers, consisting of a \textit{parametrized quantum part for exploring the search space}, and a \textit{classical part optimizing the parameters}. Our approach is the first contribution to use gate-based quantum computers for tackling query optimization. While the classical part of hybrid classical-quantum algorithms is typically difficult due to the high-dimensional parameter space, we implemented a recently suggested strategy as remedy, thus improving the solution quality. 

As current-day gate-based quantum computers are restricted in the amount of qubits and fault-tolerance, our gate-based algorithm can currently not directly compete against implementations on a quantum annealing architecture, for which quantum devices with more capacity exist. However, when comparing our hybrid approach with a competing pure quantum-annealing approach, we found that our approach has two advantages. Firstly, it can \textit{utilize the quantum device's qubits more efficiently}, as we have a direct mapping from the mathematical formulation to the quantum implementation. In contrast, an ideal mapping with quantum annealers is an NP-complete problem. Secondly, our algorithm is based on an approach that was previously shown to solve "hard" problems that quantum annealers may not be able to solve and that may be found within the MQO context.

Finally, we analytically sketched the scaling of our algorithm and found that with larger problem sizes, the computational time grows polynomially while the solution space grows exponentially. Further, the space requirements only grow linearly with the problem size.

We believe that our paper lays a solid groundwork for using hybrid classical-quantum algorithms to tackle the special problem of database optimizations as well as the generic class of binary optimization problems. 


\input{parts/bibliography} 

\end{document}

%% file: parts/bibliography.tex
\bibliographystyle{plain}
\bibliography{doc}